\documentclass[twocolumn]{aastex631}

\usepackage{multirow}
\usepackage{amsmath} % For \bm
\usepackage{bm}      % For better bold symbols
\usepackage{amsfonts} % For using \mathbb and others
\usepackage{subfigure}
\usepackage{graphicx}

\submitjournal{ApJ}

\begin{document}

\title{First X-ray polarimetric view of a Low-Luminosity Active Galactic Nucleus: the case of NGC 2110}

\correspondingauthor{Sudip Chakraborty}
\email{schakraborty2@usra.edu}

% Tier 1a - contributed in analysis 

\author[0000-0003-4872-8159]{Sudip Chakraborty}
\affiliation{Science and Technology Institute, Universities Space and Research Association, Huntsville, AL 35805, USA}
%email: schakraborty2@usra.edu

\author[0000-0003-0411-4243]{Ajay Ratheesh}
\affiliation{INAF Istituto di Astrofisica e Planetologia Spaziali, Via del Fosso del Cavaliere 100, I-00133 Roma, Italy}
%email: ajay.ratheesh@inaf.it

\author[0000-0003-3745-0112]{Daniele Tagliacozzo}
\affiliation{Dipartimento di Matematica e Fisica, Universit\`a degli Studi Roma Tre, via della Vasca Navale 84, 00146 Roma, Italy}
%email: daniele.tagliacozzo@uniroma3.it

\author[0000-0002-3638-0637]{Philip Kaaret}
\affiliation{NASA Marshall Space Flight Center, Huntsville, AL 35812, USA}
%email: philip.kaaret@nasa.gov

\author[0000-0001-5418-291X]{Jakub Podgorn{\'y}}
\affiliation{Astronomical Institute of the Czech Academy of Sciences, Bo\v{c}n\'{i} II 1401, CZ-14131 Prague, Czech Republic}
%email: jakub.podgorny@asu.cas.cz

\author[0000-0003-4952-0835]{Fr\'{e}d\'{e}ric Marin}
\affiliation{Universit\'e de Strasbourg, CNRS, Observatoire Astronomique de Strasbourg, UMR 7550, 11 rue de l'universit\'e, 67000 Strasbourg, France}
%email: frederic.marin@astro.unistra.fr

% Tier 1b - participated in discussion/editing/analysis/interpretation

\author[0000-0002-6562-8654]{Francesco Tombesi}
\affiliation{Physics Department, Tor Vergata University of Rome, Via della Ricerca Scientifica 1, 00133 Rome, Italy}
\affiliation{INAF – Astronomical Observatory of Rome, Via Frascati 33, 00040 Monte Porzio Catone, Italy}
\affiliation{INFN - Rome Tor Vergata, Via della Ricerca Scientifica 1, 00133 Rome, Italy }
%email: francesco.tombesi@roma2.infn.it

\author[0000-0003-4420-2838]{Steven R. Ehlert}
\affiliation{NASA Marshall Space Flight Center, Huntsville, AL 35812, USA}
%email:steven.r.ehlert@nasa.gov

\author[0000-0002-4945-5079]{Chien-Ting J. Chen}
\affiliation{Science and Technology Institute, Universities Space and Research Association, Huntsville, AL 35805, USA}
%email: ct.chen@nasa.gov

\author[0000-0001-5717-3736]{Dawoon E. Kim}
\affiliation{INAF Istituto di Astrofisica e Planetologia Spaziali, Via del Fosso del Cavaliere 100, I-00133 Roma, Italy}
%email:dawoon.kim@inaf.it

\author[0000-0001-9200-4006]{Ioannis Liodakis}
\affiliation{Institute of Astrophysics, Foundation for Research and Technology-Hellas, GR-70013 Heraklion, Greece}
%email:yannis.liodakis@gmail.com

\author[0000-0001-9442-7897]{Francesco Ursini}
\affiliation{Dipartimento di Matematica e Fisica, Universit\`a degli Studi Roma Tre, via della Vasca Navale 84, 00146 Roma, Italy}
%email: francesco.ursini2@uniroma3.it

% Tier 2

\author[0000-0001-9815-9092]{Riccardo Middei}
\affiliation{Space Science Data Center, Agenzia Spaziale Italiana, Via del Politecnico snc, 00133 Roma, Italy}
\affiliation{INAF – Astronomical Observatory of Rome, Via Frascati 33, 00040 Monte Porzio Catone, Italy}
%email:riccardo.middei@ssdc.asi.it

\author[0000-0003-0331-3259]{Alessandro Di Marco}
\affiliation{INAF Istituto di Astrofisica e Planetologia Spaziali, Via del Fosso del Cavaliere 100, I-00133 Roma, Italy}
%email:alessandro.dimarco@inaf.it

\author[0000-0001-8916-4156]{Fabio La Monaca}
\affiliation{INAF Istituto di Astrofisica e Planetologia Spaziali, Via del Fosso del Cavaliere 100, I-00133 Roma, Italy}
\affiliation{Physics Department, Tor Vergata University of Rome, Via della Ricerca Scientifica 1, 00133 Rome, Italy}
%\affiliation{Dipartimento di Fisica, Universit\`{a} degli Studi di Roma ``Tor Vergata'', Via della Ricerca Scientifica 1, I-00133 Roma, Italy}
%email: fabio.lamonaca@inaf.it

\author[0000-0002-6051-6928]{Srimanta Banerjee}
%\affiliation{Nicolaus Copernicus Astronomical Center, Polish Academy of Sciences, Bartycka 18, PL-00-716 Warszawa, Poland}
\affiliation{Inter-University Centre for Astronomy and Astrophysics (IUCAA), Ganeshkhind, Pune 411007, India}
%email: srimanta.banerjee4@gmail.com

\author[0000-0001-5709-7606]{Keigo Fukumura}
\affiliation{James Madison University, 800 South Main Street, Harrisonburg, Virginia 22807, USA}
%email: fukumukx@jmu.edu

\author[0000-0002-2203-7889]{W. Peter Maksym}
\affiliation{NASA Marshall Space Flight Center, Huntsville, AL 35812, USA}
%email: walter.p.maksym@nasa.gov

\author[0000-0001-7374-843X]{Romana Miku{\v{s}}incov{\'a}}
\affiliation{INAF Istituto di Astrofisica e Planetologia Spaziali, Via del Fosso del Cavaliere 100, I-00133 Roma, Italy}
%email: romana.mikusincova@uniroma3.it

\author[0000-0003-3956-0331]{Rodrigo Nemmen}
\affiliation{Universidade de S\~ao Paulo, Instituto de Astronomia, Geof\'{\i}sica e Ci\^encias Atmosf\'ericas, Departamento de Astronomia, S\~ao Paulo, SP 05508-090, Brazil}
\affiliation{Kavli Institute for Particle Astrophysics and Cosmology (KIPAC), Stanford University, Stanford, CA 94305, USA}
%email: rodrigo.nemmen@iag.usp.br

\author[0000-0001-6061-3480]{Pierre-Olivier Petrucci}
\affiliation{Universit\'e Grenoble Alpes, CNRS, IPAG, 38000 Grenoble, France}
%email: pierre-olivier.petrucci@univ-grenoble-alpes.fr

\author[0000-0002-7781-4104]{Paolo Soffitta}
\affiliation{INAF Istituto di Astrofisica e Planetologia Spaziali, Via del Fosso del Cavaliere 100, I-00133 Roma, Italy}
%email: paolo.soffitta@inaf.it

\author[0000-0003-2931-0742]{Ji\v{r}\'{i} Svoboda}
\affiliation{Astronomical Institute of the Czech Academy of Sciences, Bo\v{c}n\'{i} II 1401, CZ-14131 Prague, Czech Republic}
%email: jiri.svoboda@asu.cas.cz



\begin{abstract}

Low-Luminosity Active Galactic Nuclei (LLAGN) provides a unique view of Comptonization and non-thermal emission from accreting black holes in the low-accretion rate regime. However, to decipher the exact nature of the Comptonizing corona in LLAGN, its geometry and emission mechanism must be understood beyond the limits of spectro-timing techniques. Spectro-polarimetry offers the potential to break the degeneracies between different coronal emission models. Compton-thin LLAGN provide an opportunity for such spectro-polarimetric exploration in the 2-8 keV energy range using IXPE. In this work, we carry out a spectro-polarimetric analysis of the first IXPE observation, in synergy with a contemporaneous NuSTAR observation, of an LLAGN: NGC 2110. Using 554.4 ks of IXPE data from October 2024, we constrain the 99\% upper limit on the Polarization Degree (PD) to be less than 8.3\% assuming the corresponding Polarization Angle (PA) to be aligned with the radio jet, and less than 3.6\% if in the perpendicular direction. In the absence of a significant PD detection, the PA remains formally unconstrained, yet the polarization significance contours appear to be aligned with the radio jet, tentatively supporting models in which the corona is radially extended in the plane of the disk. We also carry out detailed Monte Carlo simulations using \textsc{monk} and \textsc{STOKES} codes to test different coronal models against our results and compare the polarization properties between NGC 2110 and brighter Seyferts. 

\end{abstract}



\section{Introduction} \label{sec:intro}

Accretion flows surrounding accreting black holes are thought to consist of hot, geometrically thick corona Comptonizing lower energy seed photons originating from cooler, geometrically thin accretion disk. In the X-ray spectrum, this gives rise to the dominant power-law component, along with a high-energy cut-off determined by the temperature of the corona \citep{Brenneman_2014}. X-ray spectra of accreting black holes often contain additional features: an iron line complex around 6.4 keV, an occasional ``Compton hump" around 30 keV, both originating from reflection of the Comptonized photons off the disk. 
The disk-corona systems of both stellar mass black holes in Galactic black hole X-ray binaries (BHBs) and supermassive black holes (SMBHs) in active galactic nuclei (AGN) undergo evolution in terms of the overall accretion rate, as well as the properties and relative contribution of the accretion disk and the corona. In BHBs, this manifests as the `q'-diagram \citep{Fender_2004,Remillard_2006}. The AGN population also occupies a similar parameter space \citep{Kording_2006,Diaz_2023}.
Accreting black holes, across several orders of magnitude in the black hole masses and accretion rates, also display a remarkable similarity in their Comptonization properties, manifesting as a geometry-independent strong anti-correlation between the electron temperature ($kT_{\rm e}$) and the optical depth ($\tau$) of the coronae \citep{Chakraborty_2023}.

The geometry of the corona can have a significant impact on Comptonization processes and on the ionization structure of the inner accretion flow \citep{Poutanen_2018}. However, X-ray spectroscopy is often insufficient in differentiating between the coronal geometries on its own, due to the myriad of degeneracies in Comptonization and reflection features.
The launch of Imaging X-Ray Polarimetry Explorer \cite[IXPE; ][]{Weisskopf_2022} has demonstrated how invaluable polarization is to break these degeneracies and to distinguish the coronal geometries of BHBs
\citep{Krawczynski_2022} and luminous radio-quiet AGN (E.g., NGC 4151 \citep{Gianolli_2023}, MCG-05-23-16 \citep{Tagliacozzo_2023, Marinucci_2022}, IC4329A \citep{Ingram_2023}) alike. 

Low-luminosity active galactic nuclei (LLAGN) are low intrinsic luminosity (bolometric luminosities $L_{\rm bol}\sim10^{38-43}$ erg s$^{-1}$) and low-accretion (significantly sub-Eddington, with Eddington ratio $\lambda_{\rm Edd} \sim 10^{-2}-10^{-5}$ \citep{Ho_2009}) AGN, possibly constituting a notable evolutionary stage for SMBHs \citep{Shin_2010}.
Despite exhibiting power-law dominated X-ray spectra akin to their higher-accretion rate counterparts, LLAGN diverge from brighter AGN in several key aspects. Notably, their broadband spectral energy distributions lack the characteristic UV ``big blue bump", suggesting a potential absence of a standard, optically thick, geometrically thin accretion disk \citep{Nemmen_2006}. Instead, accretion in LLAGN is believed to be governed by radiatively inefficient accretion flows (RIAFs), which are a subset of Advection-Dominated Accretion Flows or ADAFs \citep{Narayan_1995}. Furthermore, LLAGN typically display weak or undetectable narrow Fe ${\rm K\alpha}$ emission lines \citep{Terashima_2002}.
The power-law X-ray emission in LLAGN may originate from various mechanisms, including Comptonization of synchrotron or thermal bremsstrahlung photons, or even soft blackbody radiation from a truncated accretion disk beyond $\gtrsim 100GM/c^2$ \citep{Nemmen_2014}. Analogous to the low/hard states of BHBs in the BHB `q'-diagram, LLAGN typically occupy a comparable region in the $\lambda_{\rm Edd}$ vs. disk fraction parameter space, in contrast to the soft state analogue commonly seen in higher-luminosity Seyfert galaxies \citep{Kording_2006,Fernandez_2023}.

The coronal geometry in LLAGN remains unknown and ill-explored. Additionally, although LLAGN follow the sakTme electron temperature vs optical depth anti-correlation as its brighter AGN counterparts, the derived lack of correlation between the $\tau$ and mass accretion rate in LLAGN, stands at odds with the predictions from ADAF model. Therefore, a polarimetric exploration of LLAGN is long overdue. Compared to their brighter counterparts, LLAGN provide certain distinct advantages for polarimetry:

(a) Their 2-8 keV X-ray spectra are comprised of a simple power-law, thereby reducing degeneracy between polarization from different emission processes/regions. That apart, unpolarized thermal disc emission is absent from LLAGN spectra. Therefore, 2-8 keV spectro-polarimetry of LLAGN should give us an uncontaminated view of the Comptonization region.

(b) As for reflection, the Fe k$\alpha$ emission are rather weakly polarized, whereas the Compton hump could in principle be highly polarized \citep{Matt_1993}, albeit having rather a minimal effect in the IXPE bandpass \citep{Marin_2018}. Compton-thin LLAGN are relatively unobscured and typically contain negligible reflection features, so the observed polarization should not bee too much altered by those features.

NGC 2110 (RA= 88.0474 degs ; Dec= -7.45622 degs) is a nearby (redshift $z=0.007$) S0 galaxy \citep{Vaucouleurs_1991} with a low accretion rate ($\log_{10} \lambda_{\rm Edd}=-3.67$) Seyfert 2 AGN harboring a massive black hole ($\log(M/M_{\odot})=9.38$, \cite{Diaz_2023}). NGC 2110 is thought to possess a mid/high inclination of 42-65$^{\circ}$ \citep{Arriba_2023}. The broadband X-ray spectrum of NGC 2110 is well described by an absorbed power-law with a neutral hydrogen column density of $N_{\rm H} \sim 4 \times 10^{22} \ \rm{cm^{-2}}$, a spectral index of $\Gamma \sim$1.65 \citep{Ursini_2019}, and a high-energy cut-off. Fitting with a Comptonization model yields an optical depth of $\sim 2$ and $kT_{\rm e} \sim 75$ keV \citep{Ursini_2019}. The X-ray spectrum also contains a narrow (equivalent width $\sim$20 eV) Fe K$\alpha$ line. However, it does not exhibit strong Compton reflection features, such as a "Compton hump," suggesting that the reflecting matter is Compton-thin \citep{Marinucci_2015}. NGC 2110 exhibits significant infrared emission, suggesting the presence of an extended dusty wind or a Compton-thin torus, which may account for the Compton-thin reflection \citep{Gandhi_2009,Honig_2013,Marinucci_2015}. Further, the optical polarized light spectrum of NGC 2110 has a double-peaked broad H$\alpha$ emission line indicating a hidden broad line region shaped like a disk \citep{Moran_2007ApJ}. The polarization angle in optical is almost perpendicular to arc-second scale radio structures, indicating scattering along the direction parallel to the axis of the surrounding dust torus \citep{Gonzalez_2002ApJ}.

In this paper, we report the polarimetric results from the first IXPE observation of an LLAGN, NGC 2110. In section \ref{sec:data_reduction}, we describe the IXPE and other simultaneous observations and the data reduction procedures followed. We present our analysis and results in section \ref{sec:analysis} and discuss about the implications of our results in section \ref{sec:discussion}.

\section{Spectro-polarimetric observation of NGC 2110} \label{sec:data_reduction}

The first X-ray polarization observations of NGC 2110 using IXPE was complimented with simultaneous broad X-ray band observations using Nuclear Spectroscopic Telescope Array \cite[NuSTAR; ][]{Harrison_NuSTAR_2013}. The details of both IXPE and NuSTAR data are given in the subsequent subsections. 

\subsection{IXPE} \label{sec:ixpe}

IXPE contains 3 X-Ray Telescopes, each consisting of a mirror module assembly \cite[MMA; ][]{Ramsey_2022} and a detector unit (DU). Each of these DUs, in turn, contains a gas pixel detector (GPD) which is polarization-sensitive. The tracks of photoelectrons produced by incident X-rays in the GPD are imaged to measure the linear polarization between 2-8 keV \citep{Costa_2001,Soffitta_2021}.

NGC 2110 was observed by IXPE as a part of the GO cycle 1 (ObsID 03008799, PI: Chakraborty) in two snapshots: first on 16-22 October, 2024 and then on 26-30 October, 2024, for a total exposure time of 554.4 ks. 
To reduce the instrumental background, we first process the IXPE level 2 photon event list through the rejection algorithm of \citet{DiMarco_2023}\footnote{\url{https://github.com/aledimarco/IXPE-background}}. Just before the onset of the second IXPE snapshot, there was a prominent and long-duration solar flare which affected the IXPE data. We carried out a detailed investigation of the effects of the solar flare in the spectra and polarization signatures, and systematically filtered the flares out (refer to Appendix \ref{sec:solar_flare} for details on the effects of the solar flare on the data and the filtering strategy adopted).
Upon the solar flare filtering, the total live time for DU1, DU2, and DU3, are 538.1 ks, 537.2 ks and 537.3 ks, respectively. 

After filtering, we extract the source and background counts from a circular region with 60'' radius and a concentric annulus with radii between 150'' and 300'', respectively.
This combination of source and background region selection has been demonstrated to optimize the signal-to-background ratio for weak sources \citep{DiMarco_2023}. 
We use the \textit{ftools} command \verb|ixpepolarization|, in conjunction with this source region, to assess the Minimum Detectable Polarization at 99\% confidence ($\rm{MDP_{99}}$), and find it to be 6.78\%. We then extract the source and background spectra using \textsc{XSELECT}. Finally, we create the vignetting and aperture-corrected ARF (for Stokes I) and MRF (for Stokes Q and U) files using the \textit{ftools} task \verb|ixpecalcarf|. Throughout this paper, we use unweighted analysis (weights=NONE). We also use the appropriate response matrices for GPD (version 20240125) and MMA (version 20231201).
We find the source spectrum to be consistently above the background spectrum (scaled for the region sizes) throughout the 2-8 keV band. Although we report the results in this paper using tools from HEASoft version 6.33.2, we find consistent results with \textit{ixpeobssim} \citep{Baldini_2022} as well.

\begin{figure}
\centering
	\includegraphics[width=\columnwidth]{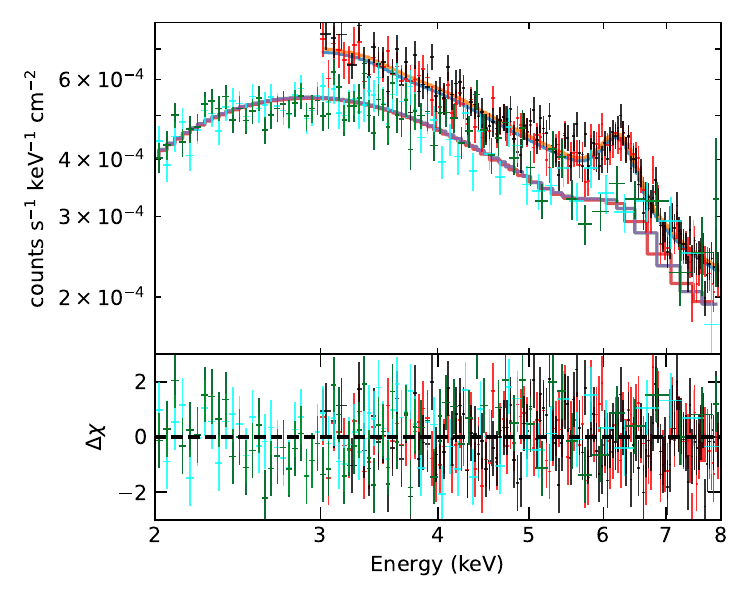}
    \caption{Contemporaneous NuSTAR (FPMA in red and FPMB in black) and IXPE (DU1 in blue, DU2 in cyan and DU3 in green), fitted with a combination of an absorbed power-law and a Gaussian. The fit is carried out in 2-8 keV energy range for IXPE and 3-8 keV energy range for NuSTAR. The Gaussian component is statistically unimportant in the standalone 2-8 keV IXPE spectra. }
    \label{fig:spec}
\end{figure}

\begin{figure}
\centering
	\includegraphics[width=\columnwidth]{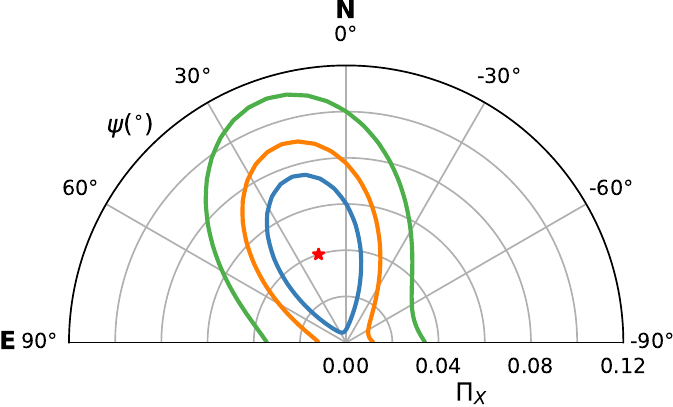}
    \caption{2-8 keV IXPE PA-PD contour plot from the best-fit spectro-polarimetric model on the joint I, Q and U spectra from all DUs in XSPEC. The radial axis is the PD ($\Pi$) and the polar axis is the PA ($\Psi$). $0^{\circ}$ is along the North and $90^{\circ}$ is along the East. The three contours are at 68.3\% (blue),90\% (orange) and 99\% (green) levels. The best-fit solution from XSPEC is denoted by a red star, with the connector to the origin displayed to guide to the corresponding PA. The shaded region denotes the $\rm{MDP_{99}}$. See Sec. \ref{sec:analysis} for a detailed description of the best-fit model. }
    \label{fig:pol_contours}
\end{figure}

\begin{figure}
\centering
	\includegraphics[width=\columnwidth]{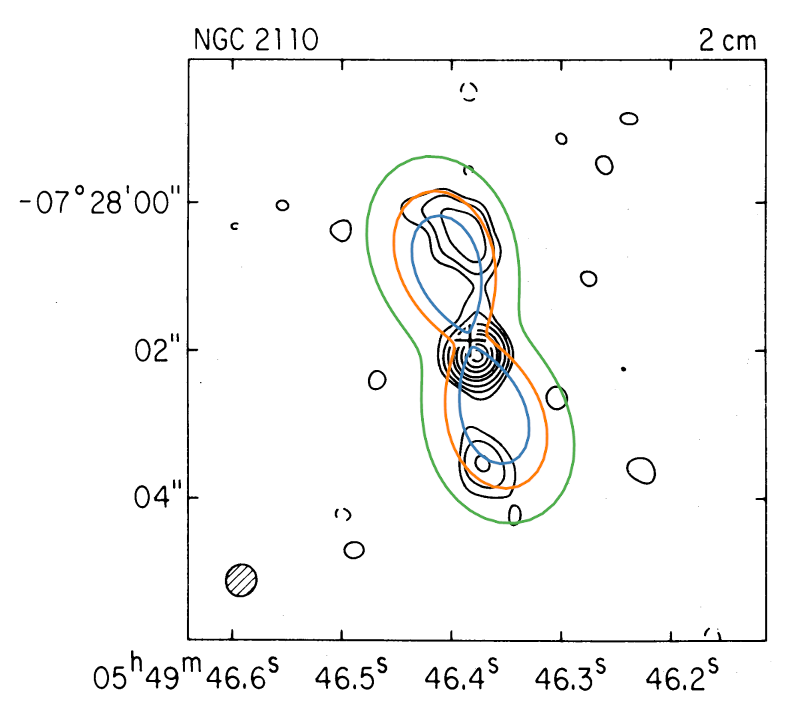}
    \caption{PA-PD contours overplotted on the 2~cm radio image taken by \citet{Ulvestad1984} to highlight the possible parallel orientation between the X-ray PA and jet.}
    \label{fig:radio}
\end{figure}

\subsection{NuSTAR} \label{sec:nustar}

NuSTAR carried out a contemporaneous observation of NGC 2110 during 19-21 October 2024 for a net exposure time of 53.4 ks in FPMA and 52.9 ks in FPMB, as a part of GO cycle 10. We use the most recent NuSTAR calibration database (CALDB) v. 2024-11-26 and process the data using the standard \verb|nupipeline| routine. We find that the standard routine sufficiently removes the aforementioned solar flare from the background light curve while creating the Good Time Intervals (GTI). Afterwards, we select circular source regions of 120'' radii to extract source and background counts, respectively. Then we compute the spectra of both source and background regions using the \verb|nuproducts| routine, and bin the spectral files consistently with a minimum of 25 counts per bin using \verb|ftgrouppha|.

\section{Results} \label{sec:analysis}

\subsection{Spectro-polarimetric Analysis} \label{sec:spectropolarimetry}

We first jointly fit the two NuSTAR FPM spectra in \textsc{XSPEC} \citep{Arnaud_1996} using an absorbed cut-off power-law in conjunction with a Gaussian to account for the iron line. We also used a constant to account for the cross-normalization between FPMA and FPMB, set at 1 for FPMA and left to vary for FPMB. We also restrict the fit within 3-50 keV since at larger energies, the background exceeds the source count rates. We also use a neutral hydrogen column density of $4\times10^{22} \ \rm{cm^{-2}}$ following \citet{Marinucci_2015}. This \verb|const|*\verb|TBabs|*\verb|(cutoffpl+gaussian)| model results in a reasonable fit with a $\chi^2$ of 974 for 1020 degrees of freedom (\textit{dof}), with the best-fit power-law photon index ($\Gamma$) of $1.60\pm0.03$, a cut-off energy $E_{\rm cut}$ of $185_{-69}^{+257}$ keV, Gaussian central energy of $6.26\pm0.02$ keV and a width of $0.27\pm0.03$ keV (at 1-$\sigma$ confidence level). The cross-normalization is found to be 1.006 for FPMB. These results are consistent with archival values and demonstrate that even though NGC 2110 has demonstrated flux variability over the years, the broadband spectrum has remained consistent. 

To arrive at the best spectral model for IXPE, we then conduct a joint spectral fit between NuSTAR and IXPE (all DUs, I spectra). To minimize the effect of the Gaussian line, we restrict both fitting ranges to 5.5 keV. As IXPE extends to lower energies than NuSTAR, it could provide a better handle over the neutral hydrogen column density ($N_{\rm H}$), while NuSTAR provides us a better estimate of the power-law photon index. With this in mind, we set up the joint fit between 2-5.5 keV IXPE and 3-5.5 keV NuSTAR with a \verb|const|*\verb|TBabs|*\verb|powerlaw| model in XSPEC, where we fix the cross normalization constant to 1 for NuSTAR FPMA and set to vary for all others, tie the NuSTAR $N_{\rm H}$ to IXPE and tie the IXPE $\Gamma$ to the NuSTAR value. The resulting fit yields a $\chi^2$ of 410 for 377 \textit{dof}, with a best-fit $N_{\rm H}$ of $(4.8\pm0.2)\times10^{22} \ \rm{cm^{-2}}$ and a $\Gamma$ of $1.47\pm0.06$ (at 1-$\sigma$ confidence level). The cross-normalization constants between NuSTAR FPMA and IXPE are found to be 0.76, 0.76 and 0.74 for DU1, DU2 and DU3, respectively. Finally, we use these $N_{\rm H}$ and $\Gamma$ values to fit the full 2-8 keV spectra for all IXPE DUs with a \verb|const|*\verb|TBabs|*\verb|powerlaw| model (cross-normalization constant fixed at 1 for DU1 and free to vary for the other DUs), which yields a best-fit $\chi^2$ of 304 for 257 \textit{dof}. The 2-8 keV flux is found to be $1.7\times10^{-11} \ \rm{erg \ cm^{-2} \ s^{-1}}$. Addition of a Gaussian component to this model provides no improvement in the fit statistic, with the resultant \textit{f-test} probability of 0.6. Therefore, we use this simple absorbed power-law model with $N_{\rm H}$ of $4.8\times10^{22} \ \rm{cm^{-2}}$ and $\Gamma$ of 1.47 for further spectro-polarimetric analysis.

Thus, we fit the I,Q and U spectra from all IXPE DUs simultaneously with \verb|const|*\verb|TBabs|*\verb|polconst|*\verb|powerlaw|, with the previously discussed spectral parameter values and assuming an energy-independent polarization signal. The best-fit Polarization degree (PD) is found to be $\Pi<10.1$\% at 99\% confidence level\footnote{\url{https://heasarc.gsfc.nasa.gov/docs/ixpe/analysis/IXPE_Stats-Advice.pdf}},
(see Fig. \ref{fig:pol_contours}). 
Although our results are consistent with an upper limit, and thus the PA is formally unconstrained as well, we can still compare the overall PD-PA contous to the position angle of the radio structure in NGC~2110 in order to infer the possible alignement with the radio jet and infer the geometry of the X-ray corona. 
\citet{Ulvestad1984} used the Karl G. Jansky Very Large Array (VLA) to map the radio structure of NGC~2110 at 2, 6 and 20~cm and reported a radio extension along the position angles 0--20$^\circ$, depending on the projected distance to the core, see Fig. \ref{fig:radio}. This radio position angle is consistent with being aligned with the X-ray PD-PA contour from IXPE. The alignment of the radio position angle and the X-ray PD-PA contours favors models in which the corona is radially extended in the plane of the disk \citep{Gianolli_2023}.

Since we find the PD-PA contour aligned with the radio jet, we can try to project the PD along two orthogonal axes, parallel and perpendicular to the PA, and find their respective bounds. As we are decomposing the 2-dimensional contour into two 1-dimensional contours in $\Delta \chi^2$ space, the individual limits would be lower than the overall 10.1\% upper limit. We use the best-fit PA and rotate the Stokes Q and U parameters in the level 2 event file by that amount to align them with $0^{\circ}$. The corresponding transformation between the initial \textit{(q,u)} and aligned \textit{(q',u')} are: $q'=\frac{qq_0+uu_0}{2}, \ u'=\frac{uq_0-qu_0}{2}$, with $q_0=2\cos(2\psi), \ u_0=2\sin(2\psi)$. This Stokes parameter alignment can transform the original two-dimensional problem into a one-dimensional problem, with PD being the only polarization parameter now. With the Stokes-aligned level 2 event file, we carry out the same data reduction procedure in Sec. \ref{sec:data_reduction} and fit with 
the same \verb|const|*\verb|TBabs|*\verb|polconst|*\verb|powerlaw|, this time freezing the PA to $0^{\circ}$. We find the best-fit PD to be $\Pi<8.3$\% at a 99\% confidence level and $\Pi<6.1$\% at a 90\% confidence level. Similarly, aligning the Stokes parameters to $90^{\circ}$, we find the corresponding 99\% and 90\% upper limit on PD to be $\Pi<$3.6\% and $\Pi<$1.8\%, respectively. The stricter constraint on the PD along $\Psi=90^{\circ}$, once again, indicates that the IXPE data favors an PA parallel to the jet.

\subsection{\textsc{monk} simulations} \label{sec:simulations}
We performed detailed Monte Carlo simulations to analyze the expected polarization properties from the corona of NGC 2110. For this, we utilized the radiative transfer code \textsc{monk} \citep{Zhang_2019}, which models the spectrum and polarization of Comptonized radiation emitted by a corona illuminated by a standard accretion disk. The seed optical and UV photons are derived based on the Novikov-Thorne disk emissivity, exhibiting polarization that varies from zero (when viewed face-on) to $11.7\%$ (when viewed edge-on), consistent with a semi-infinite, plane-parallel, scattering-dominated atmosphere \citep{1960ratr.book.....C}. Once emitted from the disk, the photons travel along geodesic trajectories in Kerr spacetime around the black hole, potentially reaching a distant observer, being redirected back to the disk, or disappearing through the event horizon. Photons entering the corona may undergo Compton up-scattering, which not only results in a Comptonized spectrum but also modifies their polarization characteristics. These scatterings significantly influence the radiation emitted by the disk in two primary ways: first, by altering the original spectrum through inverse Compton scattering, which elevates the photon energy (primarily shifting from optical/UV to X-ray wavelengths), thereby producing a harder spectrum; and second, by changing both the intensity and alignment of the overall polarization signature at higher energies. The Stokes parameters of the scattered photons are computed in the electron rest frame before being converted to the observer's frame (Boyer-Lindquist). By counting the photons that successfully reach the observer, we generate a flux and polarization spectrum. Since scattering results in linearly polarized photons, the code specifically calculates the Stokes parameters $Q$ and $U$, while the parameter $V$ is set to zero.

We tested four coronal models: the ``spherical lamppost'', consisting in a spherical source (of radius $R_c$) placed at a certain height ($H$) on the spin axis of the black hole \citep{Ursini_2022}; the ``conical outflow'', characterized by a distance ($d$) from the black hole, a thickness ($t$), an opening ($\theta$), an outflowing velocity ($v$) and which is commonly associated with an aborted jet (\citealt{hen1997A&A...326...87H, conerefId0}); the ``slab corona'' (\citealt{liang1979ApJ...231L.111L, haardt_maraschi_corona1991ApJ...380L..51H}), featured by an inner and an outer radius ($R_{in}$ and $R_{out}$) and a height above the disk plane ($h$) and can occur when magnetic loops extend well above the disk plane and release energy through reconnection (e. g. \citealt{belo2017ApJ...850..141B}); the ``wedge corona'', consisting in a homogeneous cloud of electrons characterized by an opening ($\alpha$), an inner radius ($R_{in}$) that extends down to the Innermost Stable Circular Orbit (ISCO) around the central black hole and an outer radius ($R_{out}$) (\citealt{Esin_1997, esin1998ApJ...505..854E, sh2010ApJ...712..908S, Poutanen_2018}). In Fig. \ref{fig:models} we show sketches of the aforementioned coronal models.

For each of these models we built a few different configurations of geometrical parameters and black hole spins, as summarized in Tab. \ref{monk_param}. For each different model we built two main configurations, named ``STANDARD'' and ``ADAF'': the former represents a coronal configuration mostly used for luminous radio-quiet unobscoured AGN (\citealt{Marinucci_2022, Gianolli_2023, Ingram_2023, Tagliacozzo_2023, Gianolli_2024_ngc4151b}), while the latter aims to represent LLAGN configurations by truncating the accretion disk at higher radii. Aside the parameters listed in Tab. \ref{monk_param}, we assumed: an opening $\theta=30^{\circ}$ and a velocity $v=0.3c$ (where $c$ represents the speed of light) for the conical outflow; a height above the disk plane $h=1$ $R_G$ for the slab; an opening angle $\alpha=30^{\circ}$ for the wedge. Moreover, we built a seed photon source reproducing the physical parameters of NGC 2110 (mass, Eddington ratio and coronal temperature) and for each different model configuration we fine tuned the Thomson optical depth ($\tau$), reproducing a primary continuum spectral index compatible with what has been found for this source via spectroscopical analysis (see last column in Tab. \ref{monk_param}).

\begin{figure*}[h!] 
    \centering
    \begin{subfigure}
        \centering
        \includegraphics[width=0.31\textwidth]{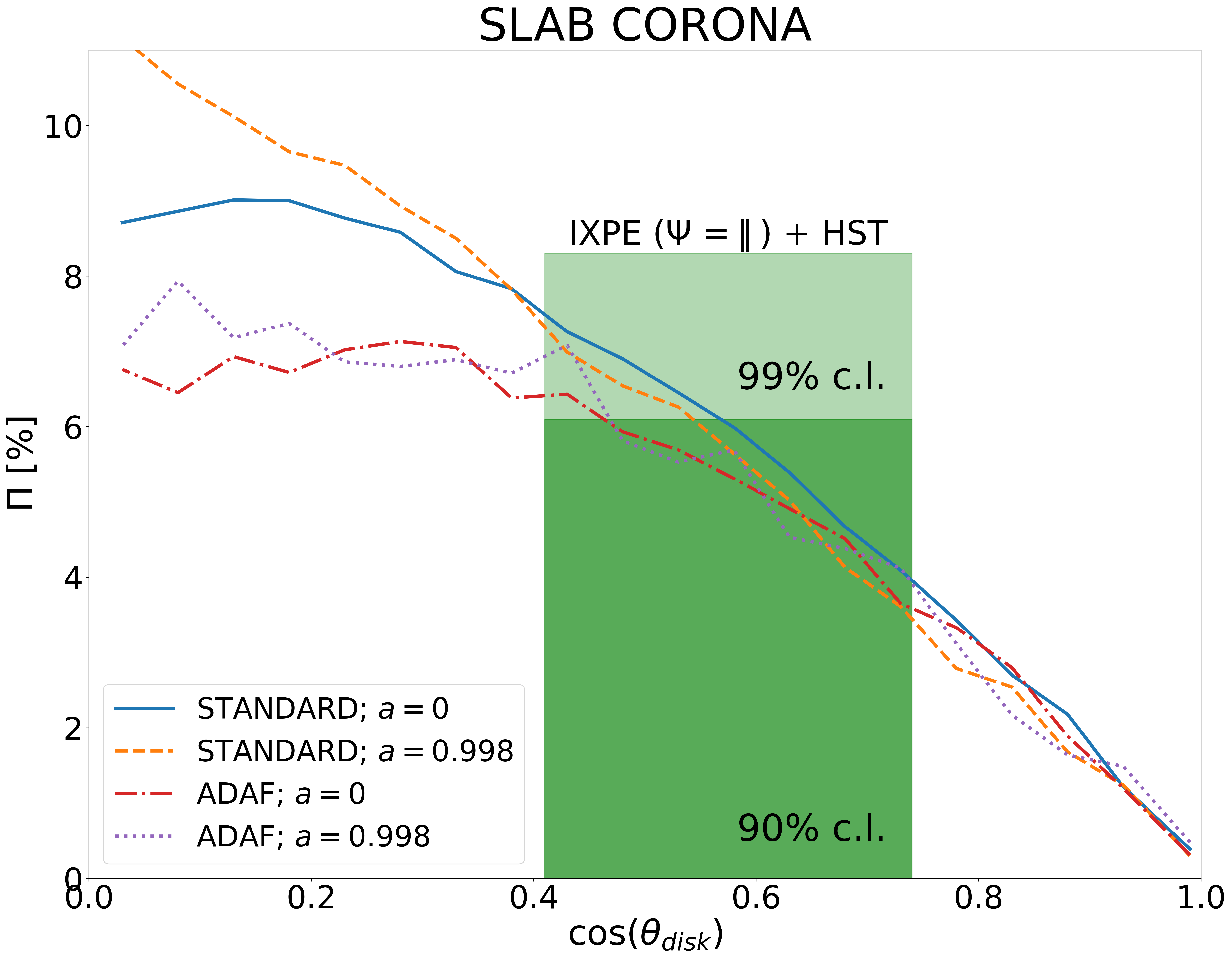}
    \end{subfigure}
    \hfill
       \begin{subfigure}
        \centering
        \includegraphics[width=0.31\textwidth]{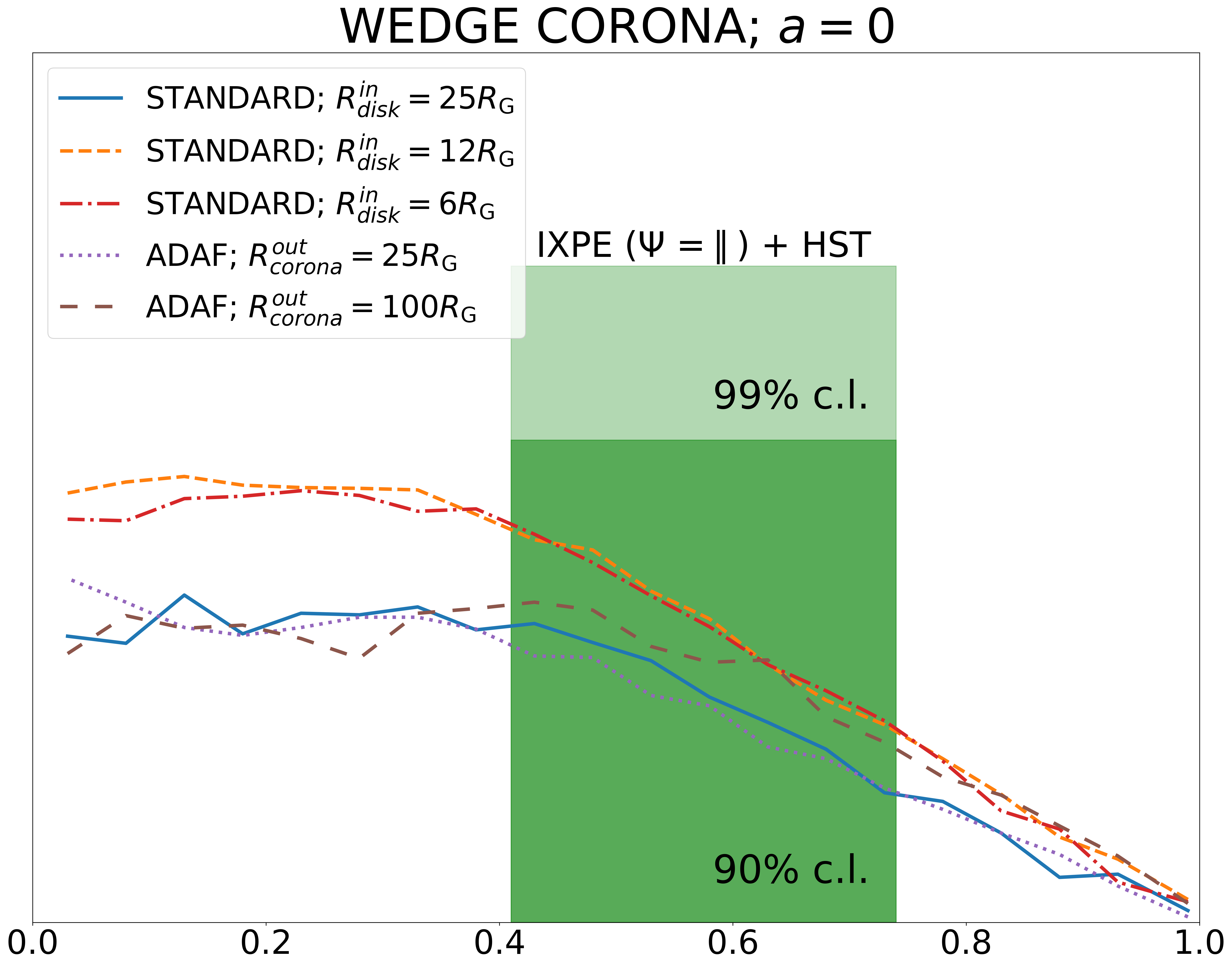}
    \end{subfigure}
    \hfill
    \begin{subfigure}
        \centering
        \includegraphics[width=0.31\textwidth]{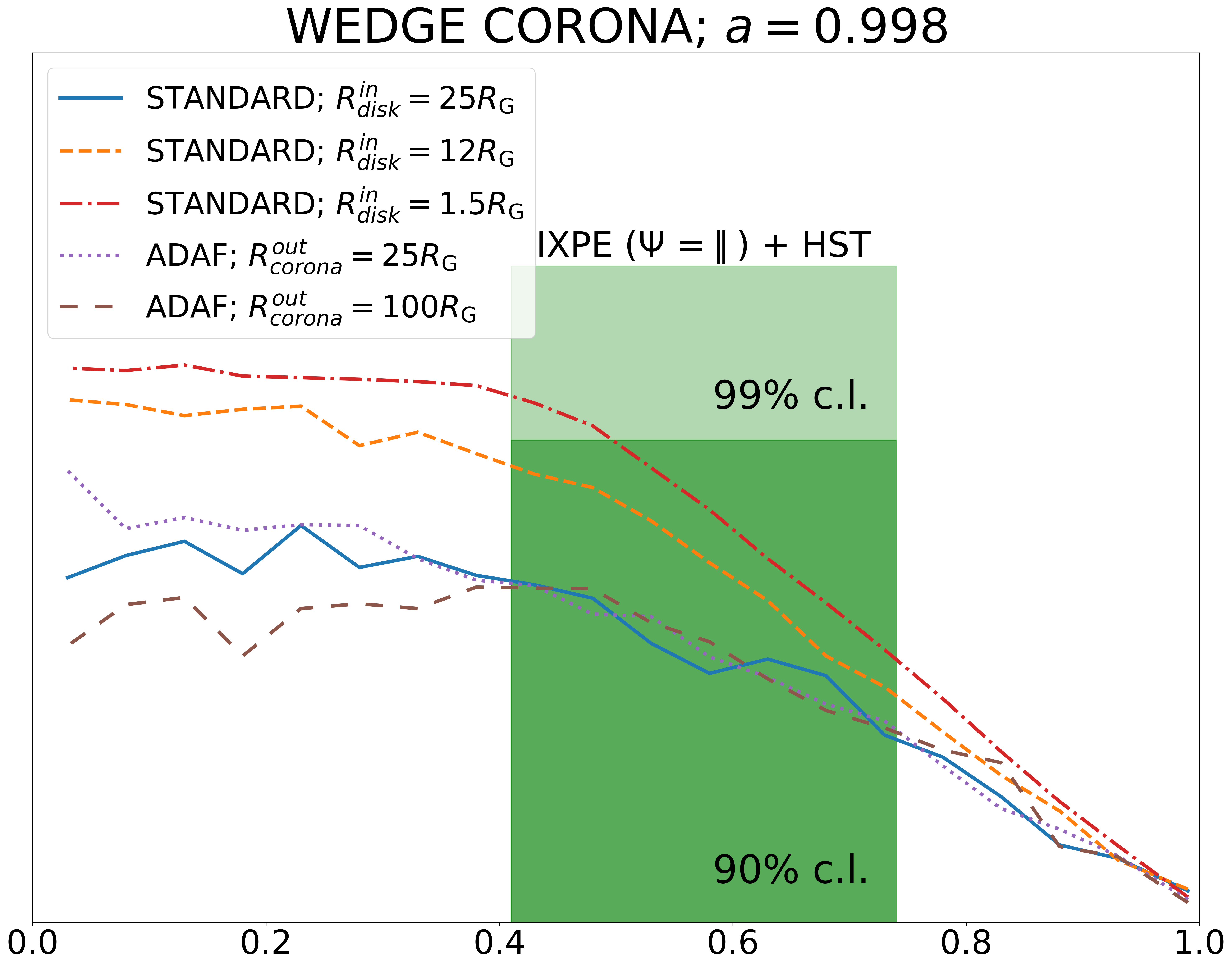}
    \end{subfigure}
    \caption{$\Pi$, summed between 2 and 8 keV, as a function of the inclination of the source ($\cos(\theta_{disk})=1$ corresponds to the face-on view) in the slab corona (\textit{left panel}) and wedge ($a=0$ in the \textit{middle panel} and $a=0.998$ in the \textit{right panel}) cases, as found via \textsc{monk} simulations. The green area represents the constraints put using IXPE data (when $\Psi$ is fixed to be parallel to the accretion disk axis $\Pi<8.3\%$ at $99\%$ c.l. and $\Pi<6.1\%$ at $90\%$ c.l.) and HST images ($42^{\circ}<\theta_{disk}<65^{\circ}$).}
    \label{monk_sim_1}
\end{figure*}

\begin{figure*}[h!] 
    \centering
    \begin{subfigure}
        \centering
        \includegraphics[width=0.417\textwidth]{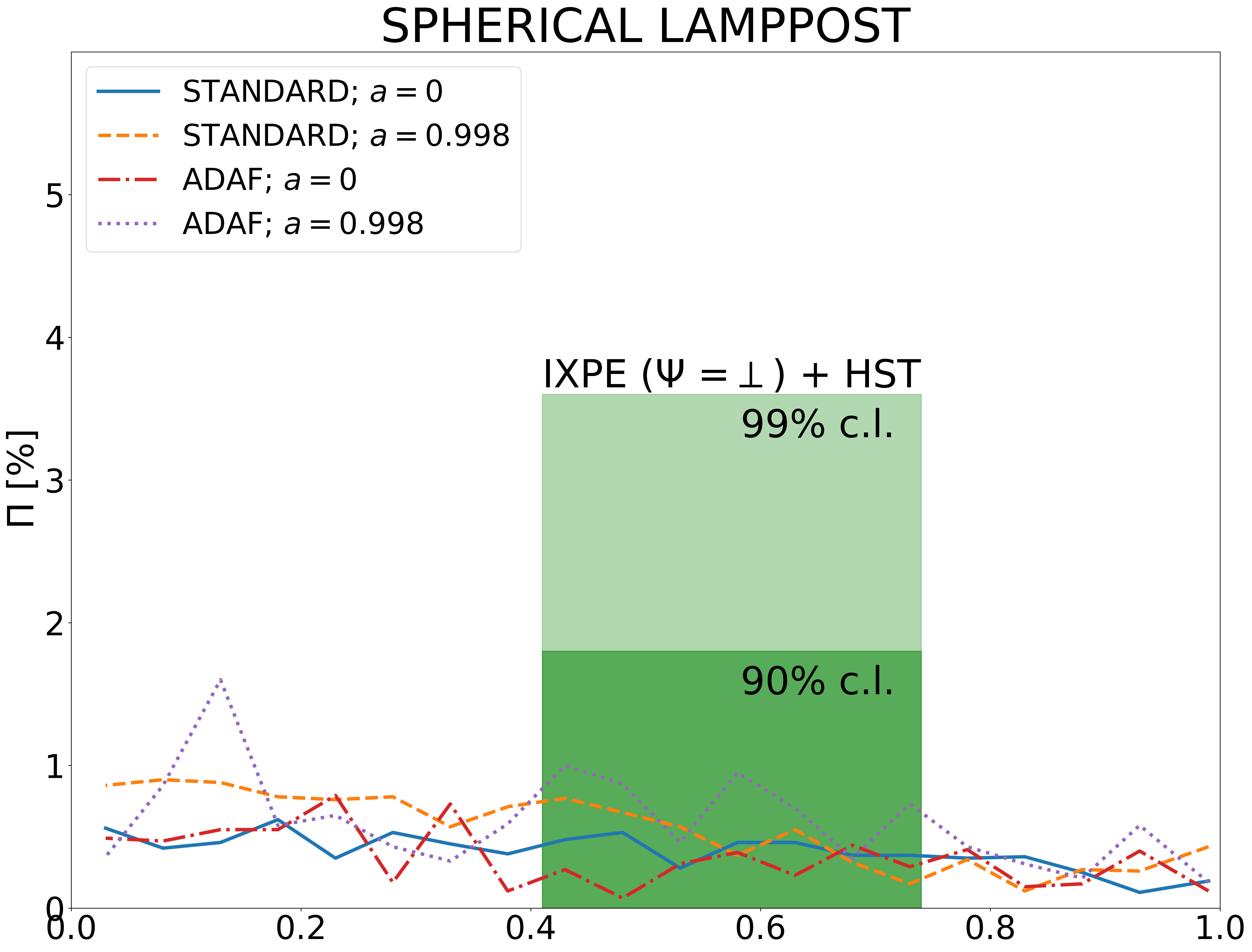}
    \end{subfigure}
    \hfill
       \begin{subfigure}
        \centering
        \includegraphics[width=0.417\textwidth]{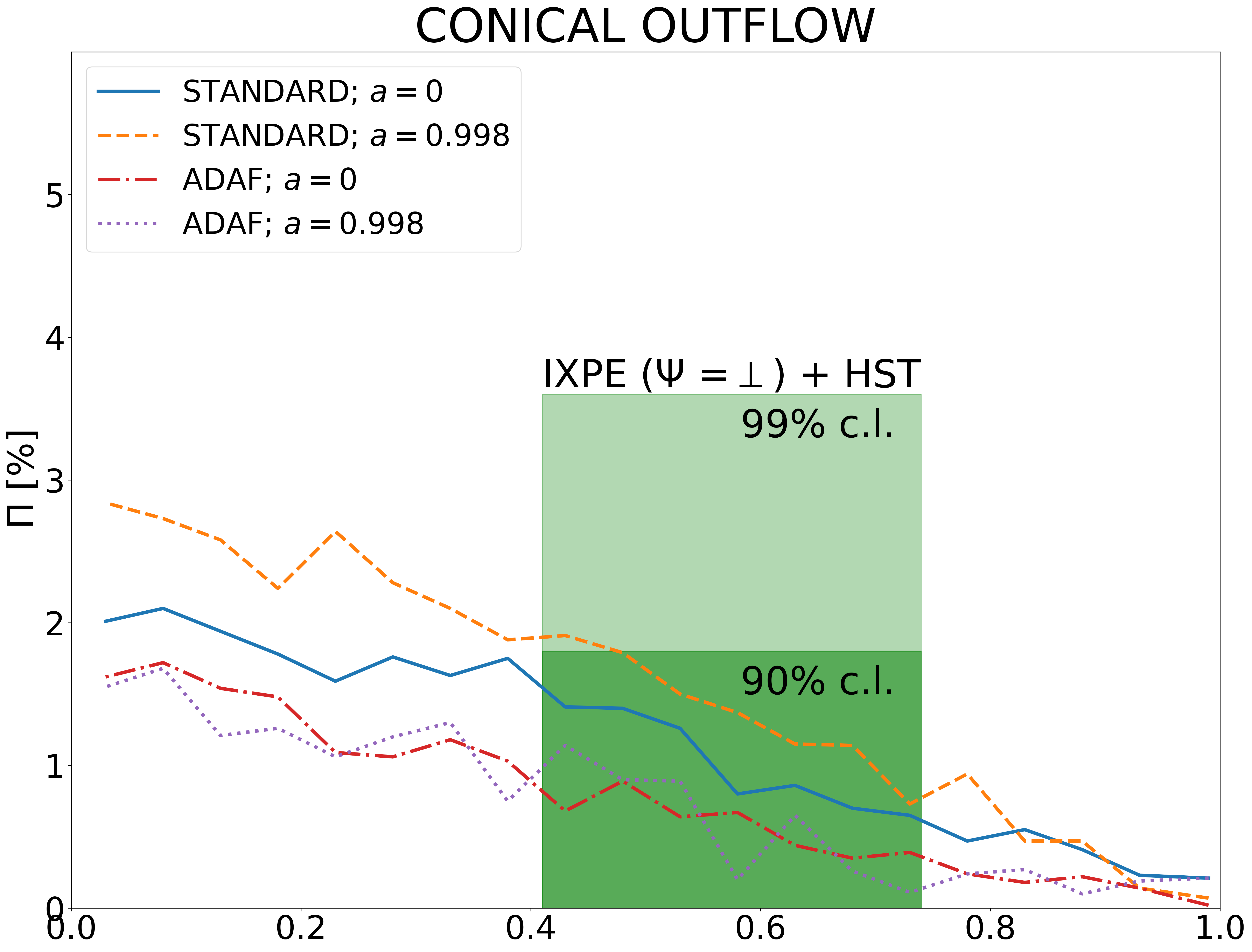}
    \end{subfigure}
        \caption{As in Fig. \ref{monk_sim_1}, but for the spherical lamppost (\textit{left panel}) and the conical outflow (\textit{right panel}). The green area represents the constraints put using HST images and IXPE data, but here with $\Psi$ fixed to be perpendicular to the accretion disk axis ($\Pi<3.6\%$ at $99\%$ c.l. and $\Pi<1.8\%$ at $90\%$ c.l.). }
    \label{monk_sim_2}
\end{figure*}

The models of the spherical lamppost and the conical outflow produce polarization angles ($\Psi$) that are perpendicular to the axis of the accretion disk, whereas the slab and wedge configurations yield angles that are parallel to it, as previously established in earlier studies (e. g. \citealt{Ursini_2022}). In Figures \ref{monk_sim_1} and \ref{monk_sim_2}, we present the polarization degree ($\Pi$) for the various models and configurations, analyzed as a function of the disk's inclination, based on the \textsc{monk} simulations. As expected, the slab corona exhibits the highest polarization fraction among the models tested, reaching $\Pi$ values of up to $12\%$. Conversely, the spherical lamppost displays the lowest polarization fraction, ranging from $0\%$ to $1\%$, due to its high symmetry. The conical outflow and wedge corona yield intermediate results, with $\Pi$ values of up to $3\%$ and $7\%$ respectively. Our findings indicate that $\Pi$ is greater when the disk is closer to the black hole. Consequently, configurations identified as ADAF, which truncate the disk at greater distances from the system's center, correspond to lower $\Pi$ values. This is primarily because the blackbody temperature of the accretion disk decreases with increasing truncation radii, necessitating more scatterings to energize the seed photons into the X-ray band. This process leads to a greater degree of randomization in the radiation and subsequently lower polarization values. Additionally, we observe that a higher spin of the black hole correlates with an increased polarization fraction. This effect arises from the combination of elevated disk temperatures which, as noted earlier, contribute to radiation de-randomization, and the increased Keplerian velocities of the corona and disk, which disrupt system symmetry more significantly at higher spin rates. 

\begin{figure*}[ht] 
    \centering
    \begin{subfigure}
        \centering
        \includegraphics[width=0.415\textwidth]{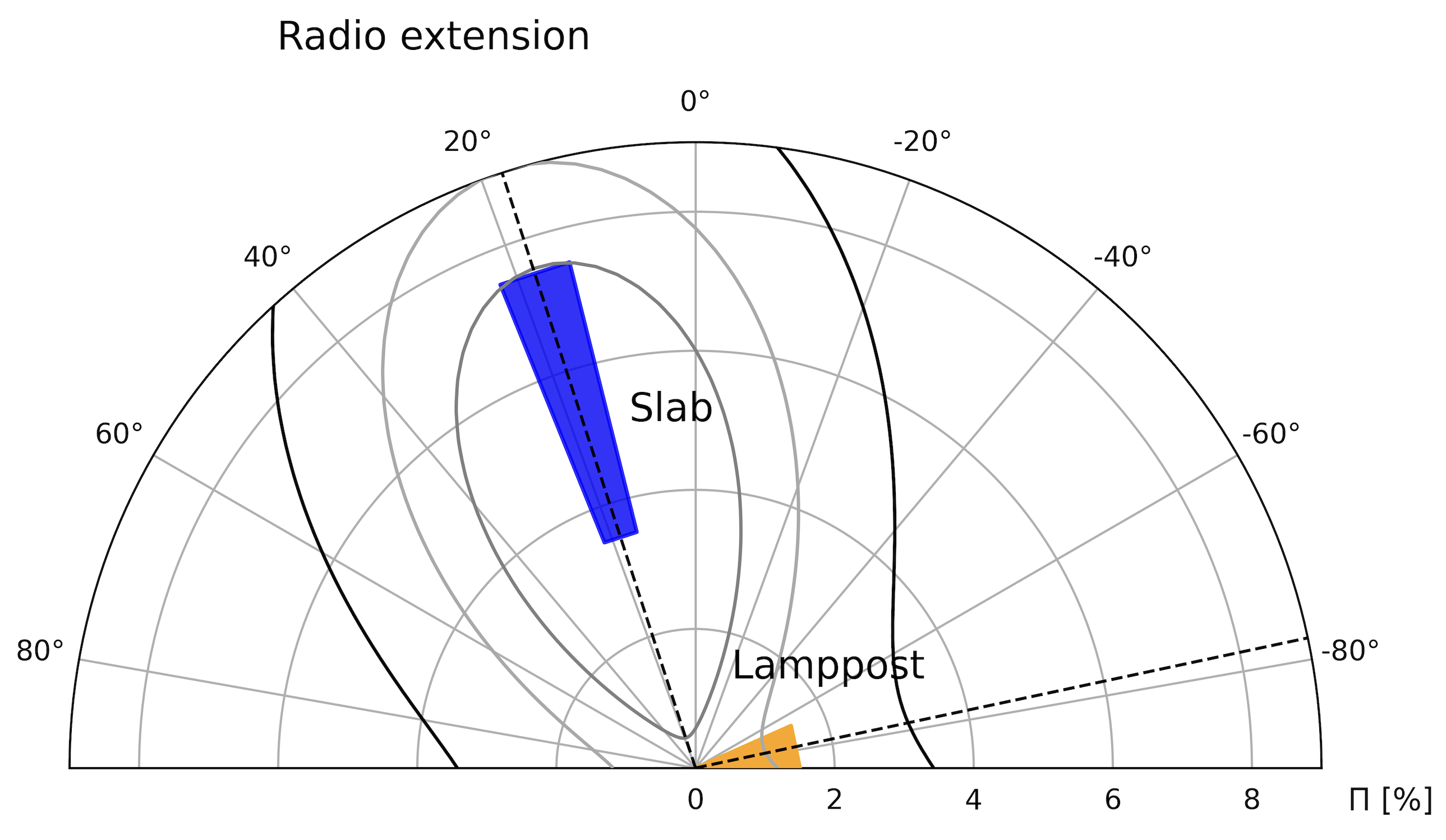}
    \end{subfigure}
    \hfill
       \begin{subfigure}
        \centering
        \includegraphics[width=0.4\textwidth]{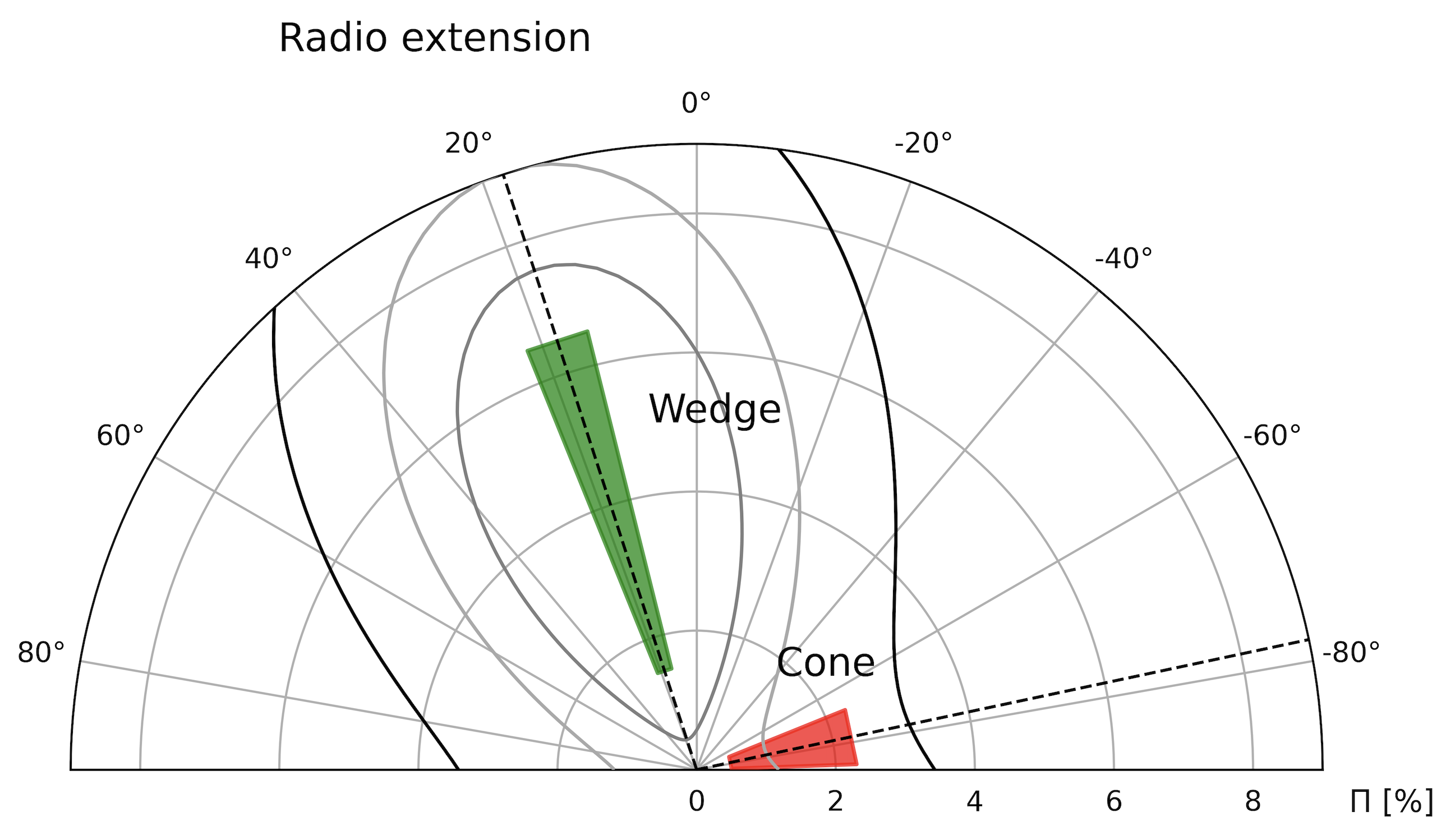}
    \end{subfigure}
    \caption{Comparison between \textsc{monk} simulations and the contour plot of the IXPE data analysis. Different coronal geometries are shown: slab (in blue) and spherical lamppost (in orange) in the \textit{left panel}, wedge (in green) and cone (in red) in the \textit{right panel}. Coloured regions of the plot represent the expected $\Pi$ for all the tested configurations in the $42^{\circ}<\theta_{disk}<65^{\circ}$ range. The black-dotted line at $18^{\circ}$ represents the supposed elongation of the radio emission, while the black-dotted line at -$78^{\circ}$ represents the direction orthogonal to the radio extension.}
    \label{polar_plots}
\end{figure*}

We also compared our simulations findings with the constraints put on the source polarization by IXPE ($\Pi<8.3\%$ when $\Psi$ is parallel to the accretion disk axis and $\Pi<3.6\%$ when $\Psi$ is perpendicular to it, at $99\%$ c.l.) and on the source inclination by HST ($42^{\circ}<\theta_{disk}<65^{\circ}$; \citealt{Gonzalez_2002ApJ}). These constraints are represented by the green regions in Figs. \ref{monk_sim_1} and \ref{monk_sim_2}. Also, in Fig. \ref{polar_plots} we show the expected polarization properties of the different models superimposed on the contour plot obtained from the analyses. Unfortunately, as we can see from all of that, we were not able to rule out any of the tested geometric configurations, due to the relatively high upper limit on the polarization fraction found within this work.

\subsection{Simulations of distant reflection} \label{sec:distant_reflection}

The \textsc{monk} simulations presented in the previous section do not take into account reprocessing in the distant, parsec-scale components of AGN. To quantitatively estimate the corresponding simulation uncertainty, i.e. the alteration of the polarization state of the X-rays from the entire nucleus by including such components, we performed additional calculations. First, we used the \textsc{xsstokes\_disc} model \citep{Podgorny_2024b} to estimate X-ray polarization due to reflection from a distant nearly neutral accretion disc and the broad line region, both assumed to reside in the equatorial plane with negligible height relative to radial extension from the center. If on the contrary the corona has non-negligible height above the equatorial plane (such as for the hot inner-accretion flows, ADAF-like coronal geometries, wedges, or coronae extended along the polar axis), it will subtend some solid angle from the point of view of the equatorial reflector. For simplicity, we integrated the reflected emission uniformly in azimuthal direction and assumed some maximum subtending angle $\delta_\textrm{i}$ measured from the normal towards the equatorial plane, which effectively represents the geometrical extension of the corona for the reflector. We refer to \cite{Podgorny_2024b} for all details on the adopted geometry of scattering and on the reprocessing spectro-polarimetric tables used. For a primary power-law emission with $\Gamma = 1.5$, in the most extreme scenario for $\cos{\delta_\textrm{i}} = 0.2$, for 7\% polarized primary emission with PA parallel to the principal axis, and for an observer inclined at $65^{\circ}$, we obtained the 2--8 keV reflected emission to be $30\%$ polarized with PA parallel to the principal axis. Any corona subtending a larger solid angle, and/or any less polarized primary emission, and/or any lower inclination, would result in a lower polarization degree induced by such reflection. Hence, if such reflection would produce at most 5\% of the total flux in 2--8 keV, it will add at most $\sim1\%$ of polarization on top of the \textsc{monk} simulation PD results presented for the wedge or slab coronae in Figure \ref{monk_sim_1}. For the wedge or slab configurations leading to unpolarized distant reflection, on the contrary, we estimate additional dilution to the expected PD of the inner emission by at most $\sim0.4\%$. For the spherical lamp-post or conical coronae, where \textsc{monk} predicts perpendicularly polarized inner emission, see Figure \ref{monk_sim_2}, distant equatorial reprocessing will generally depolarize such primary emission, or in the most extreme case cause the total 2--8 keV total emission to be at most $\sim0.3\%$ polarized parallelly with the axis.

Lastly, we estimated the impact of reprocessing of the Comptonized emission inside a purely neutral dusty torus and scattering off fully ionized polar winds or narrow line regions by performing additional simulations with the Monte Carlo code STOKES \citep{Goosmann_2007, Marin_2012, Marin_2015, Marin_2018}. We followed the exact same procedures as in \cite{Podgorny_2024a}, but we assumed a Compton-thin AGN with equatorial column density $N_\textrm{H} = 5\times 10^{22} \, \rm{cm}^{-2}$ for the torus. We chose $\tau_\textrm{wind} = 0.03$ for the winds, as the necessity of adding such highly elevated (or polar) component for interpretation of the IXPE data of the Compton-thick type-2 AGN, the Circinus Galaxy, has been shown in \cite{Podgorny_2024a} and \cite{Tanimoto_2023}. We tested a simplified primary source of emission, located in the center of the axially symmetric system, producing unpolarized power-law emission towards the poles, and polarized power-law emission towards the equator: PD = 2\% with perpendicular PA with the axis; PD = 0\%; PD = 4\%, 6\%, and 10\% with parallel PA to the axis. Assuming inclinations $65^{\circ}$, $54^{\circ}$ and $61^{\circ}$, half-opening angles of the torus $45^{\circ}$, $60^{\circ}$ and $75^{\circ}$, both measured from the polar axis, we obtained at most 0.5\% difference in the total polarization fraction in 2--8 keV with respect to the primary source emission. We conclude that any additional contribution from distant reprocessing will not significantly deviate the \textsc{monk} simulation predictions away from the PD upper limits for NGC 2110 obtained by IXPE and presented in this study.

\section{Discussions and Conclusions} \label{sec:discussion}

To summarize, we have carried out a detailed polarimetric investigation of the first LLAGN observed by IXPE, NGC 2110. We found the October 2024 IXPE observation to be affected by a solar flare, having its imprint in both the spectrum and polarization signatures. Therefore, we carefully investigated and removed the corresponding times while creating the GTI. We utilized the superior spectral capability of the simultaneous NuSTAR observation to find the spectral model that best describes the IXPE I spectra of all DUs. We used this best-fit model in XSPEC to determine the corresponding PD and PA. Furthermore, we used a rotation of Stokes Q and U parameters for the events to turn the problem into a 1D problem thereby resulting in a tighter constraint on the PD. Even though we get an upper limit instead of detection, we have a positive indication that the 2-8 keV PA is parallel to the radio jet for NGC 2110. For PA parallel to the radio jet axis, we determine the 99\% upper limit on the PD to be 8.3\%, and for PA perpendicular to the jet, we find the 99\% upper limit on the PD to be 3.6\%.

To explore the physical implications of the measured upper limits on the PD, we carried out Monte Carlo simulations with \textsc{monk} code for four different coronal models (spherical lamppost, conical outflow, slab and wedge) and two different configurations each: a ``STANDARD" one used by previous studies of luminous radio-quiet AGN, and an ``ADAF" proxy where the accretion disk is considerably truncated. We used a seed photon source representing the physical parameters of NGC and tuned the optical depth in each configuration to reproduce the X-ray spectral index consistent with our spectroscopic analysis. The spherical lamppost and conical outflow models produce PA perpendicular to the accretion disk axis, with the spherical lamppost yielding the lowest PD $\sim$0-1\% and the conical outflow leading to PD$\lesssim$3\%. On the other hand, the slab and wedge geometries produce PA parallel to the accretion disk axis, with wedge corona producing intermediate PD$\lesssim$7\% and slab corona producing the highest PD of values up to 12\%. In all the cases, the ADAF scenario results in a lower PD than the STANDARD one, possibly due to the randomization from multiple scatterings required for Comptonization of relatively low energy seed photons (from the highly truncated disk) into the 2--8 keV energy band in the ADAF scenario. The PD is also found to be generally higher for higher black hole spin values. Although the current upper limits both along the radio jet (or accretion disk) axis and orthogonal to it, are not sufficient to meaningfully rule out any of the models, certain STANDARD configurations (slab, wedge, conical outflow) predict PD close to the 90\% upper limits, especially for high spin values (Fig. \ref{monk_sim_1},\ref{monk_sim_2}). 
For a Compton-thin, LLAGN source viewed under a moderate inclination, the distant reflection has a rather small to negligible impact on observable 2--8 keV polarization properties. Therefore, the polarization predictions from the \textsc{monk} simulations carried out for the Comptonization component can be considered as representative of the observed total emission from NGC 2110.

The PA of the Compton-upscattered X-ray photons from the corona aligning with the extended radio jet has also been observed in other bright Seyfert-1 AGNs, in sources such as NGC 4151 \citep{Gianolli_2023, Gianolli_2024_ngc4151b}, IC 4329A \citep{Ingram_2023}, and MCG-05-23-16 \citep{Marinucci_2022,Tagliacozzo_2023}. A similar trend is seen in certain BHBs, including Cyg X-1 \citep{Krawczynski_2022}, Swift J1727.8-1613 \citep{Veledina_2023ApJ_Swj1727a, Ingram_2024ApJ_Swj1727b, Podgorny_2024_swj1727d}, and GX 339-4 \citep{Mastroserio_2025_gx3394}, especially in the hard state and intermediate states when the emission from the Corona dominates in the IXPE band. Hence this points to the fact that the geometry of the Comptonizing region in highly accreting AGNs and LLAGNs could be similar. An intriguing comparison can be made with the black hole binary Swift J1727.8-1613, where the coronal geometry remains similar in the hard state across two different luminosities, differing by approximately two orders of magnitude, corresponding to the top-right and bottom-right regions of the hysteresis loop in hardness intensity diagram \citep{Podgorny_2024_swj1727d}. Hence, Comptonizing geometry being similar across different luminosities could be a universal feature in black hole systems. Furthermore, the similarities of Comptonizing region in black hole binaries and AGNs indicate a universal coronal geometry across different mass scales. \cite{Chakraborty_2023} found an anti-correlation between the optical depth and the electron temperature which holds for LLAGN, brighter AGNs and BHBs, indicating a similarity in physics of Comptonization across mass scales and luminosities. The constraints on the coronal geometry NGC 2110 by IXPE in this work hence further point towards the idea that both the physics and geometry of Comptonization in black hole systems could be similar across varying blac khole masses and luminosities.

\begin{figure}
\centering
\includegraphics[width=\columnwidth]{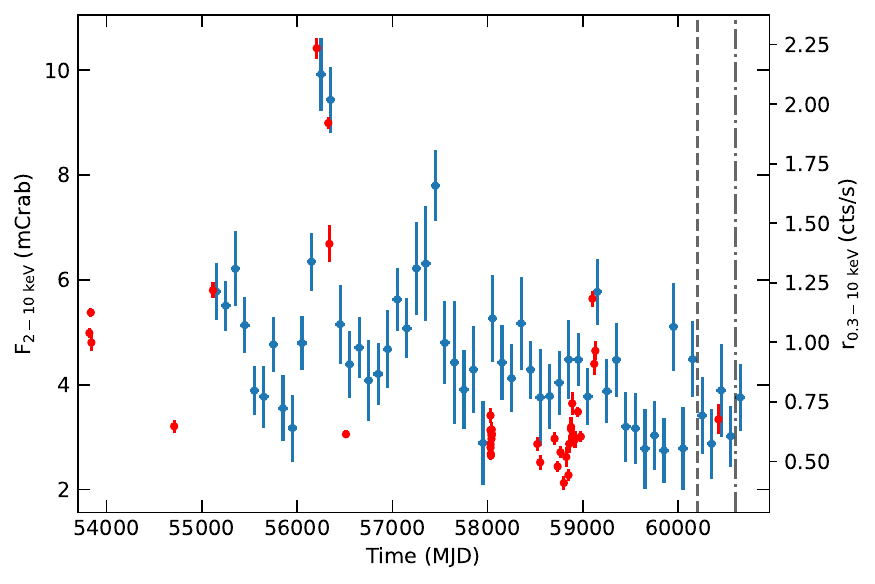}
    \caption{Long-term X-ray lightcurves for NGC 2110; 2-10 keV \textit{MAXI}/GSC with 100 days bins in blue, and 0.3-10 keV \textit{Swift}/XRT in red, plotted per ObsID. We used the daily light curves (\url{http://maxi.riken.jp/top/lc.html}) for \textit{MAXI}/GSC and automated \textit{Swift}/XRT light curves from UKSSDC (\url{https://www.swift.ac.uk/user_objects/}).}
    \label{fig:long_lightcurve}
\end{figure}

It has to be mentioned, however, that NGC 2110 diminished by at least a factor of 2 in 2-8 keV flux between the IXPE GO1 deadline and the eventual observation (as can be seen from the long-term lightcurve in Fig. \ref{fig:long_lightcurve}), thereby significantly increasing the $\rm{MDP_{99}}$ for IXPE. Despite this rapid change in flux, the spectrum has stayed relatively unchanged. Therefore, further IXPE observations should result in a lower $\rm{MDP_{99}}$ overall, and can potentially lead to a proper detection of the X-ray polarization from NGC 2110. This calls for a subsequent follow-up of NGC 2110 by IXPE. From Fig. \ref{monk_sim_1} and \ref{monk_sim_2}, we can see that a confirmed detection of PA along the jet axis will rule out conical outflow and lamppost coronae, and a PD detection even within a few \% of the 90\% upper limit could potentially rule out certain configurations of slab and wedge coronae. Future X-ray polarization observations of AGNs with IXPE and next-generation X-ray polarimeters like eXTP\citep{Santangelo2024}, across various luminosity ranges, especially the low luminosity regime would definitely improve our understanding of accretion-ejection processes in black-holes.
\\

% \begin{acknowledgments}
The Imaging X-ray Polarimetry Explorer (IXPE) is a joint US and Italian mission. The US contribution is supported by the National Aeronautics and Space Administration (NASA) and led and managed by its Marshall Space Flight Center (MSFC), with industry partner Ball Aerospace (contract NNM15AA18C). The Italian contribution is supported by the Italian Space Agency (Agenzia Spaziale Italiana, ASI) through contract ASI-OHBI-2017-12-I.0, agreements ASI-INAF-2017-12-H0 and ASI-INFN-2017.13-H0, its Space Science Data Center (SSDC), and by the Istituto Nazionale di Astrofisica (INAF) and the Istituto Nazionale di Fisica Nucleare (INFN) in Italy.
This research used data products provided by the IXPE Team (MSFC, SSDC, INAF, and INFN) and distributed with additional software tools by the High-Energy Astrophysics Science Archive Research Center (HEASARC), at NASA Goddard Space Flight Center (GSFC). This research has also made use of data from the NuSTAR mission, a project led by the California Institute of Technology, managed by the Jet Propulsion Laboratory, and funded by the National Aeronautics and Space Administration. Data analysis was performed using the NuSTAR Data Analysis Software (NuSTARDAS), jointly developed by the ASI Science Data Center (SSDC, Italy) and the California Institute of Technology (USA).
The USRA coauthors gratefully acknowledge NASA funding through contract 80NSSC24M0035. 
I.L. was funded by the European Union ERC-2022-STG - BOOTES - 101076343. J.P. and J.S. acknowledge the GACR project 21-06825X and the institutional support from RVO:67985815. Views and opinions expressed are however those of the author(s) only and do not necessarily reflect those of the European Union or the European Research Council Executive Agency. Neither the European Union nor the granting authority can be held responsible for them.

% \end{acknowledgments}


\vspace{5mm}
\facilities{IXPE, NuSTAR}

\software{astropy \citep{2013A&A...558A..33A,2018AJ....156..123A},  ixpeobssim \citep{Baldini_2022}, XSPEC \citep{Arnaud_1996}, \textsc{monk}  \citep{Zhang_2019}, \textsc{STOKES} \citep{Goosmann_2007, Marin_2012, Marin_2015, Marin_2018}}

\appendix
\setcounter{figure}{0}
\renewcommand{\thefigure}{A\arabic{figure}}
\setcounter{table}{0}
\renewcommand{\thetable}{A\arabic{table}}

\section{Effects and filtering of the solar flare}\label{sec:solar_flare}

\begin{figure}
\centering
\includegraphics[width=0.5\columnwidth]{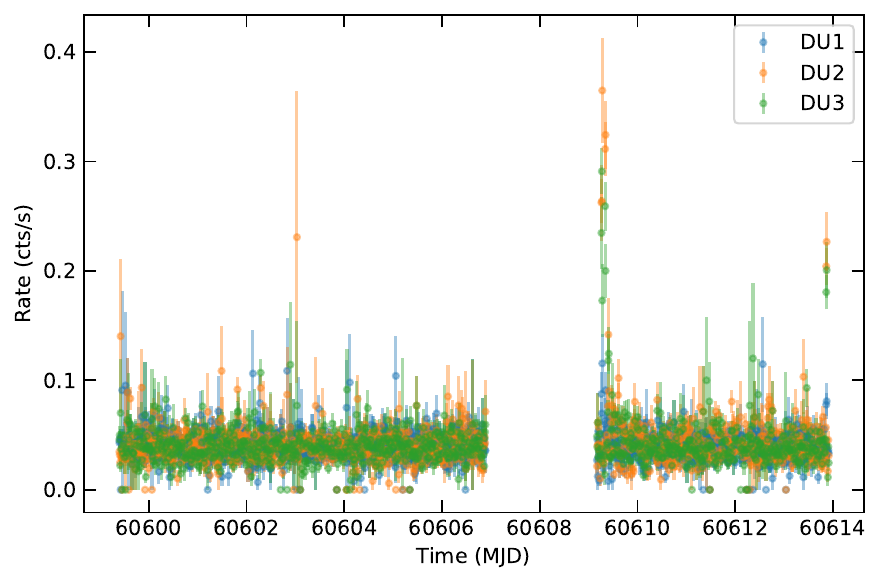}
    \caption{Binned and unweighted 2-8 keV IXPE I counting lightcurve, for all DUs. The effect of the solar flare can be noticed at the onset of the second observation segment.}
    \label{fig:lightcurve}
\end{figure}

\begin{figure}
\centering
	\includegraphics[width=0.5\columnwidth]{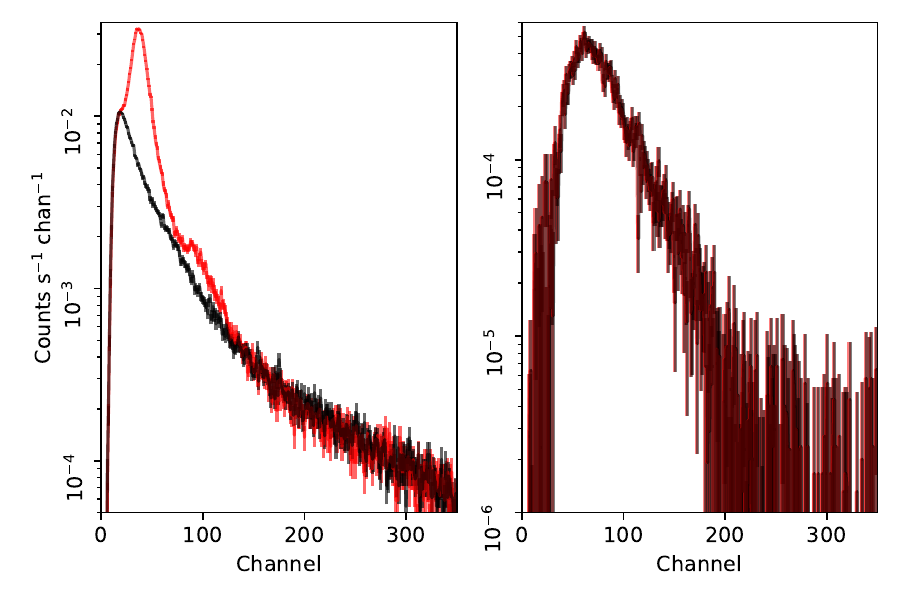}
    \caption{Effect of the solar flare in the unweighted IXPE I channel spectrum during the day (red) and night (black) for DU2. Left panel: the channel spectra for the entire field of view without any region election. The prominent effect of the Si and Al activation lines can be observed in the daytime spectrum. Right panel: the channel spectrum after source region selection and background rejection and subtraction. The effect of the solar flare is greatly minimized after the region selection and taking into account the background. The night-time spectrum has been shifted by 1 channel for better visibility.}
    \label{fig:channel_spec}
\end{figure}

\begin{figure*}
    \centering
    \begin{subfigure}
        \centering
        \includegraphics[width=0.5\columnwidth]{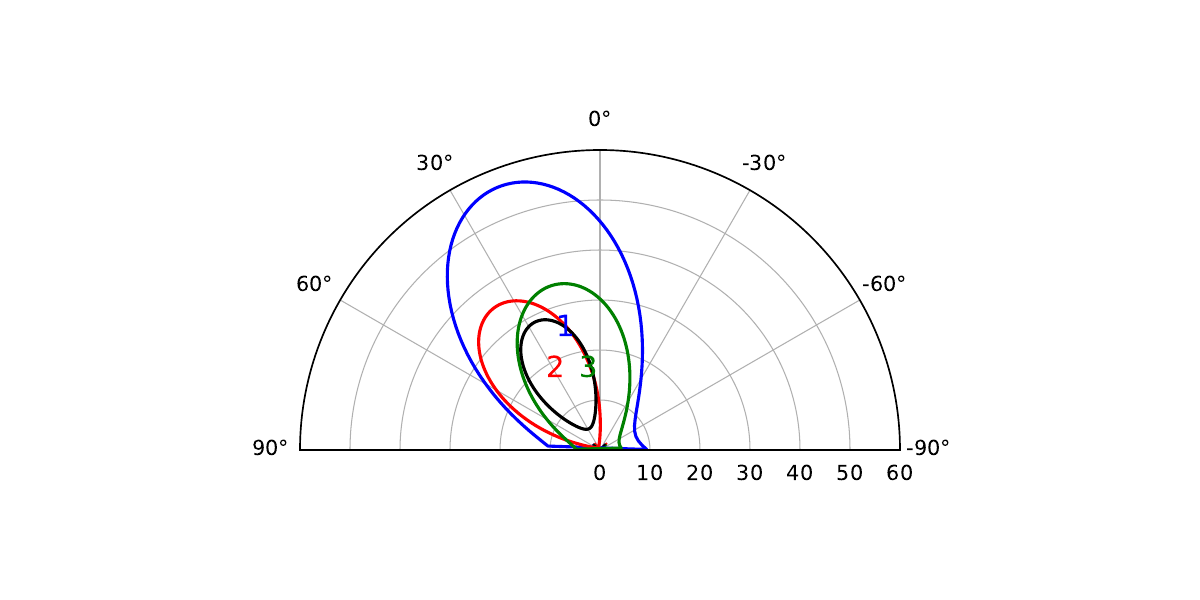}
    \end{subfigure}
    \hfill
    \begin{subfigure}
        \centering
        \includegraphics[width=0.5\columnwidth]{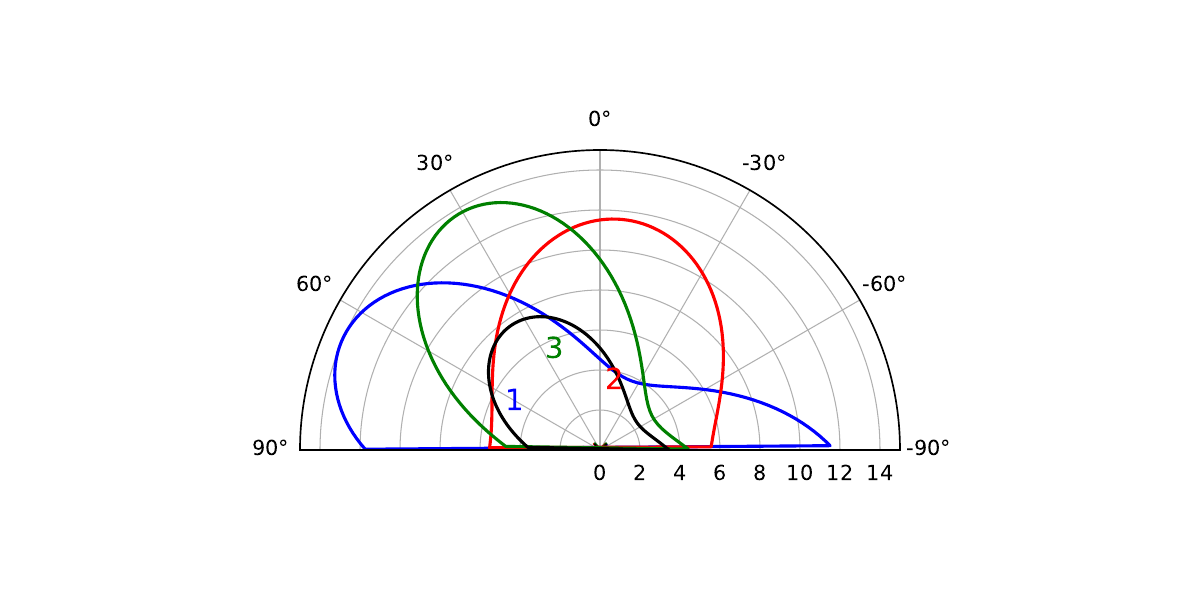}
    \end{subfigure}
    \caption{Polarization signal from unweighted events in IXPE DU1 (blue), DU2 (red), DU3 (green), as well as over all the DUs (black), arising from the solar flare in the background region described in Sec. \ref{sec:ixpe}. The contours are at 99\%. The radial axis is the Polarization Degree (PD) and the polar axis is the Polarization Angle (PA). $0^{\circ}$ is along the North and $90^{\circ}$ is along the East. Top panel: The strong 2-8 keV polarization signal in the background region during the time of the solar flare. Bottom panel: the significantly unpolarized background during the times excluding the solar flare.}
    \label{fig:flare_pol}
\end{figure*}

As mentioned in Sec. \ref{sec:ixpe}, a long-duration solar flare occured immediately before the second IXPE snapshot, the effect of which can be seen in the IXPE lightcurve in Fig. \ref{fig:lightcurve}. The effect of the solar flare was also prominent in the channel spectrum of the DUs, especially DU2 and DU3. The resulting Si and Al activation lines can be observed in the spacecraft daytime spectra (in red) as opposed to the night-time spectra (Fig. \ref{fig:channel_spec}, left panel). Upon selecting the source/background regions and doing backgorund subtraction, the day/night differences vanish (Fig. \ref{fig:channel_spec}, right panel). 
To remove the solar flares from the data, we filtered the count rates in the 2–8 keV band in all the DUs. We excised the intervals in which the count rate in any DU within a 240 s time bin exceeded the mean count rate in that DU by $\geq$ 3 times the error on the count rate. To test the effectiveness of this solar flare mitigation strategy, we investigate the polarization signal in background regions in the IXPE image during the solar flare time intervals (Fig. \ref{fig:flare_pol}, left panel) and the time intervals with the flares removed (Fig. \ref{fig:flare_pol}, right panel). From Fig. \ref{fig:flare_pol}, we can note strong polarization signal in the background region during the flare, which vanishes upon the removal of the flare.

\setcounter{figure}{0}
\renewcommand{\thefigure}{B\arabic{figure}}
\setcounter{table}{0}
\renewcommand{\thetable}{B\arabic{table}}

\section{\textsc{monk} simulation setup and parameters of the tested geometries}\label{sec:monk_setup}

\begin{figure} [h!]
\centering
	\includegraphics[width=0.9\columnwidth]{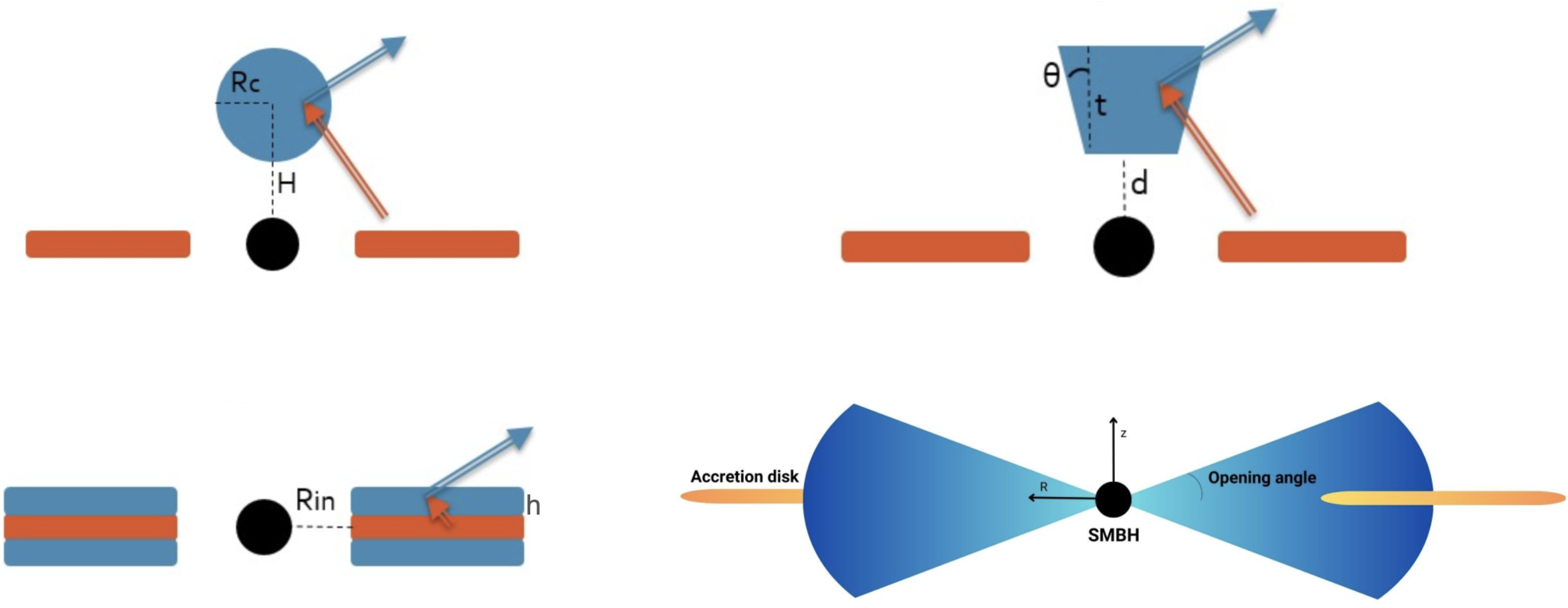}
    \caption{Coronal models tested with \textsc{monk}. \textit{Upper left}: spherical lamppost; \textit{upper right}: conical outflow; \textit{lower left}: slab corona; \textit{lower right}: wedge corona. Sketches are from \cite{Ursini_2022} and \cite{Tagliacozzo_2023}.}
    \label{fig:models}
\end{figure}

\begin{table}[h!]
\centering
\begin{tabular}{|c|c|c|c|c|c|c|c|}
\hline
\textbf{Geometry}                  & \textbf{Model}                     & \textbf{BH spin}                & $\bm{R_{disk}^{in}}$ $\bm{[R_G]}$ & $\bm{R_{disk}^{out}}$ $\bm{[R_G]}$ & \textbf{H} $\bm{[R_G]}$                 & $\bm{R_{C}}$ $\bm{[R_G]}$            & $\bm{\tau}$ \\ \hline
\multirow{4}{*}{Spherical Lamppost} & \multirow{2}{*}{STANDARD} & 0                      & 6                           & 100                          & 10                            & 7                              & 1.78   \\ \cline{3-8} 
                          &                           & 0.998                  & 1.5                         & 100                          & 5                             & 2                              & 1.66   \\ \cline{2-8} 
                          & \multirow{2}{*}{ADAF}     & 0                      & 100                         & 1000                         & 10                            & 7                              & 1.79   \\ \cline{3-8} 
                          &                           & 0.998                  & 100                         & 1000                         & 5                             & 2                              & 1.67   \\ \hline
\textbf{Geometry}                  & \textbf{Model}                     & \textbf{BH spin}                & $\bm{R_{disk}^{in}}$ $\bm{[R_G]}$ & $\bm{R_{disk}^{out}}$ $\bm{[R_G]}$ & \textbf{d} $\bm{[R_G]}$                 & \textbf{t} $\bm{[R_G]}$                  & $\bm{\tau}$ \\ \hline
\multirow{4}{*}{Conical Outflow}     & \multirow{2}{*}{STANDARD} & 0                      & 6                           & 100                          & 5                             & 15                             & 1.16   \\ \cline{3-8} 
                          &                           & 0.998                  & 1.5                         & 100                          & 3                             & 10                             & 1.06   \\ \cline{2-8} 
                          & \multirow{2}{*}{ADAF}     & 0                      & 100                         & 1000                         & 5                             & 15                             & 1.21   \\ \cline{3-8} 
                          &                           & 0.998                  & 100                         & 1000                         & 3                             & 10                             & 1.09   \\ \hline
\textbf{Geometry}                  & \textbf{Model}                     & \textbf{BH spin}                 & $\bm{R_{disk}^{in}}$ $\bm{[R_G]}$ & $\bm{R_{disk}^{out}}$ $\bm{[R_G]}$ & $\bm{R_{corona}^{in}}$ $\bm{[R_G]}$ & $\bm{R_{corona}^{out}}$ $\bm{[R_G]}$ & $\bm{\tau}$ \\ \hline
\multirow{4}{*}{Slab Corona}     & \multirow{2}{*}{STANDARD} & 0                      & 6                           & 100                          & 6                             & 100                            & 0.66   \\ \cline{3-8} 
                          &                           & 0.998                  & 1.5                         & 100                          & 1.5                           & 100                            & 0.60   \\ \cline{2-8} 
                          & \multirow{2}{*}{ADAF}     & 0                      & 100                         & 1000                         & 6                             & 100                            & 0.76   \\ \cline{3-8} 
                          &                           & 0.998                  & 100                         & 1000                         & 1.5                           & 100                            & 0.75   \\ \hline
\textbf{Geometry}                  & \textbf{Model}                     & \textbf{BH spin}                 & $\bm{R_{disk}^{in}}$ $\bm{[R_G]}$ & $\bm{R_{disk}^{out}}$ $\bm{[R_G]}$ & $\bm{R_{corona}^{in}}$ $\bm{[R_G]}$ & $\bm{R_{corona}^{out}}$ $\bm{[R_G]}$ & $\bm{\tau}$ \\ \hline
\multirow{10}{*}{Wedge Corona}   & \multirow{6}{*}{STANDARD} & \multirow{3}{*}{0}     & 25                          & 100                          & 6                             & 25                             & 2.24   \\ \cline{4-8} 
                          &                           &                        & 12                          & 100                          & 6                             & 25                             & 3.10   \\ \cline{4-8} 
                          &                           &                        & 6                           & 100                          & 6                             & 25                             & 3.50   \\ \cline{3-8} 
                          &                           & \multirow{3}{*}{0.998} & 25                          & 100                          & 1.5                           & 25                             & 2.63   \\ \cline{4-8} 
                          &                           &                        & 12                          & 100                          & 1.5                           & 25                             & 3.49   \\ \cline{4-8} 
                          &                           &                        & 1.5                         & 100                          & 1.5                           & 25                             & 4.37   \\ \cline{2-8} 
                          & \multirow{4}{*}{ADAF}     & \multirow{2}{*}{0}     & 100                         & 1000                         & 6                             & 25                             & 2.19   \\ \cline{4-8} 
                          &                           &                        & 100                         & 1000                         & 6                             & 100                            & 2.84   \\ \cline{3-8} 
                          &                           & \multirow{2}{*}{0.998} & 100                         & 1000                         & 1.5                           & 25                             & 2.53   \\ \cline{4-8} 
                          &                           &                        & 100                         & 1000                         & 1.5                           & 100                            & 2.95   \\ \hline
\end{tabular}
\caption{List of physical and geometrical parameters for the various tested coronal models within the \textsc{monk} simulations.}
\label{monk_param}
\end{table}

\bibliography{references}{}

\begin{thebibliography}{}
\expandafter\ifx\csname natexlab\endcsname\relax\def\natexlab#1{#1}\fi
\providecommand{\url}[1]{\href{#1}{#1}}
\providecommand{\dodoi}[1]{doi:~\href{http://doi.org/#1}{\nolinkurl{#1}}}
\providecommand{\doeprint}[1]{\href{http://ascl.net/#1}{\nolinkurl{http://ascl.net/#1}}}
\providecommand{\doarXiv}[1]{\href{https://arxiv.org/abs/#1}{\nolinkurl{https://arxiv.org/abs/#1}}}

\bibitem[{{Arnaud}(1996)}]{Arnaud_1996}
{Arnaud}, K.~A. 1996, Astronomical Society of the Pacific Conference Series, Vol. 101, {XSPEC: The First Ten Years}, ed. G.~H. {Jacoby} \& J.~{Barnes}, 17

\bibitem[{{Astropy Collaboration} {et~al.}(2013){Astropy Collaboration}, {Robitaille}, {Tollerud}, {Greenfield}, {Droettboom}, {Bray}, {Aldcroft}, {Davis}, {Ginsburg}, {Price-Whelan}, {Kerzendorf}, {Conley}, {Crighton}, {Barbary}, {Muna}, {Ferguson}, {Grollier}, {Parikh}, {Nair}, {Unther}, {Deil}, {Woillez}, {Conseil}, {Kramer}, {Turner}, {Singer}, {Fox}, {Weaver}, {Zabalza}, {Edwards}, {Azalee Bostroem}, {Burke}, {Casey}, {Crawford}, {Dencheva}, {Ely}, {Jenness}, {Labrie}, {Lim}, {Pierfederici}, {Pontzen}, {Ptak}, {Refsdal}, {Servillat}, \& {Streicher}}]{2013A&A...558A..33A}
{Astropy Collaboration}, {Robitaille}, T.~P., {Tollerud}, E.~J., {et~al.} 2013, \aap, 558, A33, \dodoi{10.1051/0004-6361/201322068}

\bibitem[{{Astropy Collaboration} {et~al.}(2018){Astropy Collaboration}, {Price-Whelan}, {Sip{\H{o}}cz}, {G{\"u}nther}, {Lim}, {Crawford}, {Conseil}, {Shupe}, {Craig}, {Dencheva}, {Ginsburg}, {VanderPlas}, {Bradley}, {P{\'e}rez-Su{\'a}rez}, {de Val-Borro}, {Aldcroft}, {Cruz}, {Robitaille}, {Tollerud}, {Ardelean}, {Babej}, {Bach}, {Bachetti}, {Bakanov}, {Bamford}, {Barentsen}, {Barmby}, {Baumbach}, {Berry}, {Biscani}, {Boquien}, {Bostroem}, {Bouma}, {Brammer}, {Bray}, {Breytenbach}, {Buddelmeijer}, {Burke}, {Calderone}, {Cano Rodr{\'\i}guez}, {Cara}, {Cardoso}, {Cheedella}, {Copin}, {Corrales}, {Crichton}, {D'Avella}, {Deil}, {Depagne}, {Dietrich}, {Donath}, {Droettboom}, {Earl}, {Erben}, {Fabbro}, {Ferreira}, {Finethy}, {Fox}, {Garrison}, {Gibbons}, {Goldstein}, {Gommers}, {Greco}, {Greenfield}, {Groener}, {Grollier}, {Hagen}, {Hirst}, {Homeier}, {Horton}, {Hosseinzadeh}, {Hu}, {Hunkeler}, {Ivezi{\'c}}, {Jain}, {Jenness}, {Kanarek}, {Kendrew}, {Kern}, {Kerzendorf}, {Khvalko}, {King}, {Kirkby}, {Kulkarni},
  {Kumar}, {Lee}, {Lenz}, {Littlefair}, {Ma}, {Macleod}, {Mastropietro}, {McCully}, {Montagnac}, {Morris}, {Mueller}, {Mumford}, {Muna}, {Murphy}, {Nelson}, {Nguyen}, {Ninan}, {N{\"o}the}, {Ogaz}, {Oh}, {Parejko}, {Parley}, {Pascual}, {Patil}, {Patil}, {Plunkett}, {Prochaska}, {Rastogi}, {Reddy Janga}, {Sabater}, {Sakurikar}, {Seifert}, {Sherbert}, {Sherwood-Taylor}, {Shih}, {Sick}, {Silbiger}, {Singanamalla}, {Singer}, {Sladen}, {Sooley}, {Sornarajah}, {Streicher}, {Teuben}, {Thomas}, {Tremblay}, {Turner}, {Terr{\'o}n}, {van Kerkwijk}, {de la Vega}, {Watkins}, {Weaver}, {Whitmore}, {Woillez}, {Zabalza}, \& {Astropy Contributors}}]{2018AJ....156..123A}
{Astropy Collaboration}, {Price-Whelan}, A.~M., {Sip{\H{o}}cz}, B.~M., {et~al.} 2018, \aj, 156, 123, \dodoi{10.3847/1538-3881/aabc4f}

\bibitem[{{Baldini} {et~al.}(2022){Baldini}, {Bucciantini}, {Lalla}, {Ehlert}, {Manfreda}, {Negro}, {Omodei}, {Pesce-Rollins}, {Sgr{\`o}}, \& {Silvestri}}]{Baldini_2022}
{Baldini}, L., {Bucciantini}, N., {Lalla}, N.~D., {et~al.} 2022, SoftwareX, 19, 101194, \dodoi{10.1016/j.softx.2022.101194}

\bibitem[{{Beloborodov}(2017)}]{belo2017ApJ...850..141B}
{Beloborodov}, A.~M. 2017, \apj, 850, 141, \dodoi{10.3847/1538-4357/aa8f4f}

\bibitem[{{Brenneman} {et~al.}(2014){Brenneman}, {Madejski}, {Fuerst}, {Matt}, {Elvis}, {Harrison}, {Ballantyne}, {Boggs}, {Christensen}, {Craig}, {Fabian}, {Grefenstette}, {Hailey}, {Madsen}, {Marinucci}, {Rivers}, {Stern}, {Walton}, \& {Zhang}}]{Brenneman_2014}
{Brenneman}, L.~W., {Madejski}, G., {Fuerst}, F., {et~al.} 2014, \apj, 788, 61, \dodoi{10.1088/0004-637X/788/1/61}

\bibitem[{{Chakraborty} {et~al.}(2023){Chakraborty}, {Ratheesh}, {Tombesi}, {Nemmen}, \& {Banerjee}}]{Chakraborty_2023}
{Chakraborty}, S., {Ratheesh}, A., {Tombesi}, F., {Nemmen}, R., \& {Banerjee}, S. 2023, \aap, 676, L13, \dodoi{10.1051/0004-6361/202347181}

\bibitem[{{Chandrasekhar}(1960)}]{1960ratr.book.....C}
{Chandrasekhar}, S. 1960, {Radiative transfer} (New York: Dover)

\bibitem[{{Costa} {et~al.}(2001){Costa}, {Soffitta}, {Bellazzini}, {Brez}, {Lumb}, \& {Spandre}}]{Costa_2001}
{Costa}, E., {Soffitta}, P., {Bellazzini}, R., {et~al.} 2001, \nat, 411, 662, \dodoi{10.1038/35079508}

\bibitem[{{de Vaucouleurs}(1991)}]{Vaucouleurs_1991}
{de Vaucouleurs}, G. 1991, \mnras, 249, 28P, \dodoi{10.1093/mnras/249.1.28P}

\bibitem[{{Di Marco} {et~al.}(2023){Di Marco}, {Soffitta}, {Costa}, {Ferrazzoli}, {La Monaca}, {Rankin}, {Ratheesh}, {Xie}, {Baldini}, {Del Monte}, {Ehlert}, {Fabiani}, {Kim}, {Muleri}, {O'Dell}, {Ramsey}, {Rubini}, {Sgr{\`o}}, {Silvestri}, {Tennant}, \& {Weisskopf}}]{DiMarco_2023}
{Di Marco}, A., {Soffitta}, P., {Costa}, E., {et~al.} 2023, \aj, 165, 143, \dodoi{10.3847/1538-3881/acba0f}

\bibitem[{{Diaz} {et~al.}(2023){Diaz}, {Hern{\`a}ndez-Garc{\'\i}a}, {Ar{\'e}valo}, {L{\'o}pez-Navas}, {Ricci}, {Koss}, {Gonzalez-Martin}, {Balokovi{\'c}}, {Osorio-Clavijo}, {Garc{\'\i}a}, \& {Malizia}}]{Diaz_2023}
{Diaz}, Y., {Hern{\`a}ndez-Garc{\'\i}a}, L., {Ar{\'e}valo}, P., {et~al.} 2023, \aap, 669, A114, \dodoi{10.1051/0004-6361/202244678}

\bibitem[{{Esin} {et~al.}(1997){Esin}, {McClintock}, \& {Narayan}}]{Esin_1997}
{Esin}, A.~A., {McClintock}, J.~E., \& {Narayan}, R. 1997, \apj, 489, 865, \dodoi{10.1086/304829}

\bibitem[{{Esin} {et~al.}(1998){Esin}, {Narayan}, {Cui}, {Grove}, \& {Zhang}}]{esin1998ApJ...505..854E}
{Esin}, A.~A., {Narayan}, R., {Cui}, W., {Grove}, J.~E., \& {Zhang}, S.-N. 1998, \apj, 505, 854, \dodoi{10.1086/306186}

\bibitem[{{Fender} {et~al.}(2004){Fender}, {Belloni}, \& {Gallo}}]{Fender_2004}
{Fender}, R.~P., {Belloni}, T.~M., \& {Gallo}, E. 2004, \mnras, 355, 1105, \dodoi{10.1111/j.1365-2966.2004.08384.x}

\bibitem[{{Fern{\'a}ndez-Ontiveros} {et~al.}(2023){Fern{\'a}ndez-Ontiveros}, {L{\'o}pez-L{\'o}pez}, \& {Prieto}}]{Fernandez_2023}
{Fern{\'a}ndez-Ontiveros}, J.~A., {L{\'o}pez-L{\'o}pez}, X., \& {Prieto}, A. 2023, \aap, 670, A22, \dodoi{10.1051/0004-6361/202243547}

\bibitem[{{Gandhi} {et~al.}(2009){Gandhi}, {Horst}, {Smette}, {H{\"o}nig}, {Comastri}, {Gilli}, {Vignali}, \& {Duschl}}]{Gandhi_2009}
{Gandhi}, P., {Horst}, H., {Smette}, A., {et~al.} 2009, \aap, 502, 457, \dodoi{10.1051/0004-6361/200811368}

\bibitem[{{Ghisellini} {et~al.}(2004){Ghisellini}, {Haardt}, \& {Matt}}]{conerefId0}
{Ghisellini}, G., {Haardt}, F., \& {Matt}, G. 2004, \aap, 413, 535, \dodoi{10.1051/0004-6361:20031562}

\bibitem[{{Gianolli} {et~al.}(2023){Gianolli}, {Kim}, {Bianchi}, {Ag{\'\i}s-Gonz{\'a}lez}, {Madejski}, {Marin}, {Marinucci}, {Matt}, {Middei}, {Petrucci}, {Soffitta}, {Tagliacozzo}, {Tombesi}, {Ursini}, {Barnouin}, {De Rosa}, {Di Gesu}, {Ingram}, {Loktev}, {Panagiotou}, {Podgorny}, {Poutanen}, {Puccetti}, {Ratheesh}, {Veledina}, {Zhang}, {Agudo}, {Antonelli}, {Bachetti}, {Baldini}, {Baumgartner}, {Bellazzini}, {Bongiorno}, {Bonino}, {Brez}, {Bucciantini}, {Capitanio}, {Castellano}, {Cavazzuti}, {Chen}, {Ciprini}, {Costa}, {Del Monte}, {Di Lalla}, {Di Marco}, {Donnarumma}, {Doroshenko}, {Dov{\v{c}}iak}, {Ehlert}, {Enoto}, {Evangelista}, {Fabiani}, {Ferrazzoli}, {Garc{\'\i}a}, {Gunji}, {Heyl}, {Iwakiri}, {Jorstad}, {Kaaret}, {Karas}, {Kislat}, {Kitaguchi}, {Kolodziejczak}, {Krawczynski}, {La Monaca}, {Latronico}, {Liodakis}, {Maldera}, {Manfreda}, {Marscher}, {Marshall}, {Massaro}, {Mitsuishi}, {Mizuno}, {Muleri}, {Negro}, {Ng}, {O'Dell}, {Omodei}, {Oppedisano}, {Papitto}, {Pavlov}, {Peirson}, {Perri},
  {Pesce-Rollins}, {Pilia}, {Possenti}, {Ramsey}, {Rankin}, {Roberts}, {Romani}, {Sgr{\`o}}, {Slane}, {Spandre}, {Swartz}, {Tamagawa}, {Tavecchio}, {Taverna}, {Tawara}, {Tennant}, {Thomas}, {Trois}, {Tsygankov}, {Turolla}, {Vink}, {Weisskopf}, {Wu}, {Xie}, \& {Zane}}]{Gianolli_2023}
{Gianolli}, V.~E., {Kim}, D.~E., {Bianchi}, S., {et~al.} 2023, \mnras, 523, 4468, \dodoi{10.1093/mnras/stad1697}

\bibitem[{{Gianolli} {et~al.}(2024){Gianolli}, {Bianchi}, {Kammoun}, {Gnarini}, {Marinucci}, {Ursini}, {Parra}, {Tortosa}, {De Rosa}, {Kim}, {Marin}, {Matt}, {Serafinelli}, {Soffitta}, {Tagliacozzo}, {Di Gesu}, {Done}, {Marshall}, {Middei}, {Mikusincova}, {Petrucci}, {Ravi}, {Svoboda}, \& {Tombesi}}]{Gianolli_2024_ngc4151b}
{Gianolli}, V.~E., {Bianchi}, S., {Kammoun}, E., {et~al.} 2024, \aap, 691, A29, \dodoi{10.1051/0004-6361/202451645}

\bibitem[{{Gonz{\'a}lez Delgado} {et~al.}(2002){Gonz{\'a}lez Delgado}, {Arribas}, {P{\'e}rez}, \& {Heckman}}]{Gonzalez_2002ApJ}
{Gonz{\'a}lez Delgado}, R.~M., {Arribas}, S., {P{\'e}rez}, E., \& {Heckman}, T. 2002, \apj, 579, 188, \dodoi{10.1086/342675}

\bibitem[{{Goosmann} \& {Gaskell}(2007)}]{Goosmann_2007}
{Goosmann}, R.~W., \& {Gaskell}, C.~M. 2007, \aap, 465, 129, \dodoi{10.1051/0004-6361:20053555}

\bibitem[{{Haardt} \& {Maraschi}(1991)}]{haardt_maraschi_corona1991ApJ...380L..51H}
{Haardt}, F., \& {Maraschi}, L. 1991, \apjl, 380, L51, \dodoi{10.1086/186171}

\bibitem[{{Harrison} {et~al.}(2013){Harrison}, {Craig}, {Christensen}, {Hailey}, {Zhang}, {Boggs}, {Stern}, {Cook}, {Forster}, {Giommi}, {Grefenstette}, {Kim}, {Kitaguchi}, {Koglin}, {Madsen}, {Mao}, {Miyasaka}, {Mori}, {Perri}, {Pivovaroff}, {Puccetti}, {Rana}, {Westergaard}, {Willis}, {Zoglauer}, {An}, {Bachetti}, {Barri{\`e}re}, {Bellm}, {Bhalerao}, {Brejnholt}, {Fuerst}, {Liebe}, {Markwardt}, {Nynka}, {Vogel}, {Walton}, {Wik}, {Alexander}, {Cominsky}, {Hornschemeier}, {Hornstrup}, {Kaspi}, {Madejski}, {Matt}, {Molendi}, {Smith}, {Tomsick}, {Ajello}, {Ballantyne}, {Balokovi{\'c}}, {Barret}, {Bauer}, {Blandford}, {Brandt}, {Brenneman}, {Chiang}, {Chakrabarty}, {Chenevez}, {Comastri}, {Dufour}, {Elvis}, {Fabian}, {Farrah}, {Fryer}, {Gotthelf}, {Grindlay}, {Helfand}, {Krivonos}, {Meier}, {Miller}, {Natalucci}, {Ogle}, {Ofek}, {Ptak}, {Reynolds}, {Rigby}, {Tagliaferri}, {Thorsett}, {Treister}, \& {Urry}}]{Harrison_NuSTAR_2013}
{Harrison}, F.~A., {Craig}, W.~W., {Christensen}, F.~E., {et~al.} 2013, \apj, 770, 103, \dodoi{10.1088/0004-637X/770/2/103}

\bibitem[{{Henri} \& {Petrucci}(1997)}]{hen1997A&A...326...87H}
{Henri}, G., \& {Petrucci}, P.~O. 1997, \aap, 326, 87, \dodoi{10.48550/arXiv.astro-ph/9705233}

\bibitem[{{Ho}(2009)}]{Ho_2009}
{Ho}, L.~C. 2009, \apj, 699, 626, \dodoi{10.1088/0004-637X/699/1/626}

\bibitem[{{H{\"o}nig} {et~al.}(2013){H{\"o}nig}, {Kishimoto}, {Tristram}, {Prieto}, {Gandhi}, {Asmus}, {Antonucci}, {Burtscher}, {Duschl}, \& {Weigelt}}]{Honig_2013}
{H{\"o}nig}, S.~F., {Kishimoto}, M., {Tristram}, K.~R.~W., {et~al.} 2013, \apj, 771, 87, \dodoi{10.1088/0004-637X/771/2/87}

\bibitem[{{Ingram} {et~al.}(2023){Ingram}, {Ewing}, {Marinucci}, {Tagliacozzo}, {Rosario}, {Veledina}, {Kim}, {Marin}, {Bianchi}, {Poutanen}, {Matt}, {Marshall}, {Ursini}, {De Rosa}, {Petrucci}, {Madejski}, {Barnouin}, {Gesu}, {Dov{\v{c}}iak}, {Gianolli}, {Krawczynski}, {Loktev}, {Middei}, {Podgorny}, {Puccetti}, {Ratheesh}, {Soffitta}, {Tombesi}, {Ehlert}, {Massaro}, {Agudo}, {Antonelli}, {Bachetti}, {Baldini}, {Baumgartner}, {Bellazzini}, {Bongiorno}, {Bonino}, {Brez}, {Bucciantini}, {Capitanio}, {Castellano}, {Cavazzuti}, {Chen}, {Ciprini}, {Costa}, {Del Monte}, {Lalla}, {Marco}, {Donnarumma}, {Doroshenko}, {Enoto}, {Evangelista}, {Fabiani}, {Ferrazzoli}, {Garc{\'\i}a}, {Gunji}, {Heyl}, {Iwakiri}, {Jorstad}, {Kaaret}, {Karas}, {Kislat}, {Kitaguchi}, {Kolodziejczak}, {Monaca}, {Latronico}, {Liodakis}, {Maldera}, {Manfreda}, {Marscher}, {Mitsuishi}, {Mizuno}, {Muleri}, {Negro}, {Ng}, {O'Dell}, {Omodei}, {Oppedisano}, {Papitto}, {Pavlov}, {Peirson}, {Perri}, {Pesce-Rollins}, {Pilia}, {Possenti}, {Ramsey},
  {Rankin}, {Roberts}, {Romani}, {Sgr{\`o}}, {Slane}, {Spandre}, {Swartz}, {Tamagawa}, {Tavecchio}, {Taverna}, {Tawara}, {Tennant}, {Thomas}, {Trois}, {Tsygankov}, {Turolla}, {Vink}, {Weisskopf}, {Wu}, {Xie}, \& {Zane}}]{Ingram_2023}
{Ingram}, A., {Ewing}, M., {Marinucci}, A., {et~al.} 2023, \mnras, 525, 5437, \dodoi{10.1093/mnras/stad2625}

\bibitem[{{Ingram} {et~al.}(2024){Ingram}, {Bollemeijer}, {Veledina}, {Dov{\v{c}}iak}, {Poutanen}, {Egron}, {Russell}, {Trushkin}, {Negro}, {Ratheesh}, {Capitanio}, {Connors}, {Neilsen}, {Kraus}, {Iacolina}, {Pellizzoni}, {Pilia}, {Carotenuto}, {Matt}, {Mastroserio}, {Kaaret}, {Bianchi}, {Garc{\'\i}a}, {Bachetti}, {Wu}, {Costa}, {Ewing}, {Kravtsov}, {Krawczynski}, {Loktev}, {Marinucci}, {Marra}, {Miku{\v{s}}incov{\'a}}, {Nathan}, {Parra}, {Petrucci}, {Righini}, {Soffitta}, {Steiner}, {Svoboda}, {Tombesi}, {Tugliani}, {Ursini}, {Yang}, {Zane}, {Zhang}, {Agudo}, {Antonelli}, {Baldini}, {Baumgartner}, {Bellazzini}, {Bongiorno}, {Bonino}, {Brez}, {Bucciantini}, {Castellano}, {Cavazzuti}, {Chen}, {Ciprini}, {De Rosa}, {Del Monte}, {Di Gesu}, {Di Lalla}, {Di Marco}, {Donnarumma}, {Doroshenko}, {Ehlert}, {Enoto}, {Evangelista}, {Fabiani}, {Ferrazzoli}, {Gunji}, {Hayashida}, {Heyl}, {Iwakiri}, {Jorstad}, {Karas}, {Kislat}, {Kitaguchi}, {Kolodziejczak}, {La Monaca}, {Latronico}, {Liodakis}, {Maldera}, {Manfreda},
  {Marin}, {Marscher}, {Marshall}, {Massaro}, {Mitsuishi}, {Mizuno}, {Muleri}, {Ng}, {O'Dell}, {Omodei}, {Oppedisano}, {Papitto}, {Pavlov}, {Peirson}, {Perri}, {Pesce-Rollins}, {Possenti}, {Puccetti}, {Ramsey}, {Rankin}, {Roberts}, {Romani}, {Sgr{\`o}}, {Slane}, {Spandre}, {Swartz}, {Tamagawa}, {Tavecchio}, {Taverna}, {Tawara}, {Tennant}, {Thomas}, {Trois}, {Tsygankov}, {Turolla}, {Vink}, {Weisskopf}, {Xie}, \& {IXPE Collaboration}}]{Ingram_2024ApJ_Swj1727b}
{Ingram}, A., {Bollemeijer}, N., {Veledina}, A., {et~al.} 2024, \apj, 968, 76, \dodoi{10.3847/1538-4357/ad3faf}

\bibitem[{{K{\"o}rding} {et~al.}(2006){K{\"o}rding}, {Jester}, \& {Fender}}]{Kording_2006}
{K{\"o}rding}, E.~G., {Jester}, S., \& {Fender}, R. 2006, \mnras, 372, 1366, \dodoi{10.1111/j.1365-2966.2006.10954.x}

\bibitem[{{Krawczynski} {et~al.}(2022){Krawczynski}, {Muleri}, {Dov{\v{c}}iak}, {Veledina}, {Rodriguez Cavero}, {Svoboda}, {Ingram}, {Matt}, {Garcia}, {Loktev}, {Negro}, {Poutanen}, {Kitaguchi}, {Podgorn{\'y}}, {Rankin}, {Zhang}, {Berdyugin}, {Berdyugina}, {Bianchi}, {Blinov}, {Capitanio}, {Di Lalla}, {Draghis}, {Fabiani}, {Kagitani}, {Kravtsov}, {Kiehlmann}, {Latronico}, {Lutovinov}, {Mandarakas}, {Marin}, {Marinucci}, {Miller}, {Mizuno}, {Molkov}, {Omodei}, {Petrucci}, {Ratheesh}, {Sakanoi}, {Semena}, {Skalidis}, {Soffitta}, {Tennant}, {Thalhammer}, {Tombesi}, {Weisskopf}, {Wilms}, {Zhang}, {Agudo}, {Antonelli}, {Bachetti}, {Baldini}, {Baumgartner}, {Bellazzini}, {Bongiorno}, {Bonino}, {Brez}, {Bucciantini}, {Castellano}, {Cavazzuti}, {Ciprini}, {Costa}, {De Rosa}, {Del Monte}, {Di Gesu}, {Di Marco}, {Donnarumma}, {Doroshenko}, {Ehlert}, {Enoto}, {Evangelista}, {Ferrazzoli}, {Gunji}, {Hayashida}, {Heyl}, {Iwakiri}, {Jorstad}, {Karas}, {Kolodziejczak}, {La Monaca}, {Liodakis}, {Maldera}, {Manfreda},
  {Marscher}, {Marshall}, {Mitsuishi}, {Ng}, {O{\textquoteright}Dell}, {Oppedisano}, {Papitto}, {Pavlov}, {Peirson}, {Perri}, {Pesce-Rollins}, {Pilia}, {Possenti}, {Puccetti}, {Ramsey}, {Romani}, {Sgr{\`o}}, {Slane}, {Spandre}, {Tamagawa}, {Tavecchio}, {Taverna}, {Tawara}, {Thomas}, {Trois}, {Tsygankov}, {Turolla}, {Vink}, {Wu}, {Xie}, \& {Zane}}]{Krawczynski_2022}
{Krawczynski}, H., {Muleri}, F., {Dov{\v{c}}iak}, M., {et~al.} 2022, Science, 378, 650, \dodoi{10.1126/science.add5399}

\bibitem[{{Liang}(1979)}]{liang1979ApJ...231L.111L}
{Liang}, E.~P.~T. 1979, \apjl, 231, L111, \dodoi{10.1086/183015}

\bibitem[{{Marin} {et~al.}(2018){Marin}, {Dov{\v{c}}iak}, \& {Kammoun}}]{Marin_2018}
{Marin}, F., {Dov{\v{c}}iak}, M., \& {Kammoun}, E.~S. 2018, \mnras, 478, 950, \dodoi{10.1093/mnras/sty1062}

\bibitem[{{Marin} {et~al.}(2015){Marin}, {Goosmann}, \& {Gaskell}}]{Marin_2015}
{Marin}, F., {Goosmann}, R.~W., \& {Gaskell}, C.~M. 2015, \aap, 577, A66, \dodoi{10.1051/0004-6361/201525628}

\bibitem[{{Marin} {et~al.}(2012){Marin}, {Goosmann}, {Gaskell}, {Porquet}, \& {Dov{\v{c}}iak}}]{Marin_2012}
{Marin}, F., {Goosmann}, R.~W., {Gaskell}, C.~M., {Porquet}, D., \& {Dov{\v{c}}iak}, M. 2012, \aap, 548, A121, \dodoi{10.1051/0004-6361/201219751}

\bibitem[{{Marinucci} {et~al.}(2015){Marinucci}, {Matt}, {Bianchi}, {Lu}, {Arevalo}, {Balokovi{\'c}}, {Ballantyne}, {Bauer}, {Boggs}, {Christensen}, {Craig}, {Gandhi}, {Hailey}, {Harrison}, {Puccetti}, {Rivers}, {Walton}, {Stern}, \& {Zhang}}]{Marinucci_2015}
{Marinucci}, A., {Matt}, G., {Bianchi}, S., {et~al.} 2015, \mnras, 447, 160, \dodoi{10.1093/mnras/stu2439}

\bibitem[{{Marinucci} {et~al.}(2022){Marinucci}, {Muleri}, {Dovciak}, {Bianchi}, {Marin}, {Matt}, {Ursini}, {Middei}, {Marshall}, {Baldini}, {Barnouin}, {Rodriguez}, {De Rosa}, {Di Gesu}, {Harper}, {Ingram}, {Karas}, {Krawczynski}, {Madejski}, {Panagiotou}, {Petrucci}, {Podgorny}, {Puccetti}, {Tombesi}, {Veledina}, {Zhang}, {Agudo}, {Antonelli}, {Bachetti}, {Baumgartner}, {Bellazzini}, {Bongiorno}, {Bonino}, {Brez}, {Bucciantini}, {Capitanio}, {Castellano}, {Cavazzuti}, {Ciprini}, {Costa}, {Del Monte}, {Di Lalla}, {Di Marco}, {Donnarumma}, {Doroshenko}, {Ehlert}, {Enoto}, {Evangelista}, {Fabiani}, {Ferrazzoli}, {Garcia}, {Gunji}, {Hayashida}, {Heyl}, {Iwakiri}, {Jorstad}, {Kitaguchi}, {Kolodziejczak}, {La Monaca}, {Latronico}, {Liodakis}, {Maldera}, {Manfreda}, {Marscher}, {Mitsuishi}, {Mizuno}, {Ng}, {O'Dell}, {Omodei}, {Oppedisano}, {Papitto}, {Pavlov}, {Peirson}, {Perri}, {Pesce-Rollins}, {Pilia}, {Possenti}, {Poutanen}, {Ramsey}, {Rankin}, {Ratheesh}, {Romani}, {Sgr{\v{s}}}, {Slane}, {Soffitta},
  {Spandre}, {Tamagawa}, {Tavecchio}, {Taverna}, {Tawara}, {Tennant}, {Thomas}, {Trois}, {Tsygankov}, {Turolla}, {Vink}, {Weisskopf}, {Wu}, {Xie}, \& {Zane}}]{Marinucci_2022}
{Marinucci}, A., {Muleri}, F., {Dovciak}, M., {et~al.} 2022, \mnras, 516, 5907, \dodoi{10.1093/mnras/stac2634}

\bibitem[{{Mastroserio} {et~al.}(2025){Mastroserio}, {De Marco}, {Baglio}, {Carotenuto}, {Fabiani}, {Russell}, {Capitanio}, {Cavecchi}, {Motta}, {Russell}, {Dov{\v{c}}iak}, {Del Santo}, {Alabarta}, {Ambrifi}, {Campana}, {Casella}, {Covino}, {Illiano}, {Kara}, {Lai}, {Lodato}, {Manca}, {Mariani}, {Marino}, {Miceli}, {Saikia}, {Shaw}, {Svoboda}, {Vincentelli}, \& {Wang}}]{Mastroserio_2025_gx3394}
{Mastroserio}, G., {De Marco}, B., {Baglio}, M.~C., {et~al.} 2025, \apjl, 978, L19, \dodoi{10.3847/2041-8213/ad9913}

\bibitem[{{Matt}(1993)}]{Matt_1993}
{Matt}, G. 1993, \mnras, 260, 663, \dodoi{10.1093/mnras/260.3.663}

\bibitem[{{Moran} {et~al.}(2007){Moran}, {Barth}, {Eracleous}, \& {Kay}}]{Moran_2007ApJ}
{Moran}, E.~C., {Barth}, A.~J., {Eracleous}, M., \& {Kay}, L.~E. 2007, \apjl, 668, L31, \dodoi{10.1086/522697}

\bibitem[{{Narayan} \& {Yi}(1995)}]{Narayan_1995}
{Narayan}, R., \& {Yi}, I. 1995, \apj, 452, 710, \dodoi{10.1086/176343}

\bibitem[{{Nemmen} {et~al.}(2014){Nemmen}, {Storchi-Bergmann}, \& {Eracleous}}]{Nemmen_2014}
{Nemmen}, R.~S., {Storchi-Bergmann}, T., \& {Eracleous}, M. 2014, \mnras, 438, 2804, \dodoi{10.1093/mnras/stt2388}

\bibitem[{{Nemmen} {et~al.}(2006){Nemmen}, {Storchi-Bergmann}, {Yuan}, {Eracleous}, {Terashima}, \& {Wilson}}]{Nemmen_2006}
{Nemmen}, R.~S., {Storchi-Bergmann}, T., {Yuan}, F., {et~al.} 2006, \apj, 643, 652, \dodoi{10.1086/500571}

\bibitem[{{Peralta de Arriba} {et~al.}(2023){Peralta de Arriba}, {Alonso-Herrero}, {Garc{\'\i}a-Burillo}, {Garc{\'\i}a-Bernete}, {Villar-Mart{\'\i}n}, {Garc{\'\i}a-Lorenzo}, {Davies}, {Rosario}, {H{\"o}nig}, {Levenson}, {Packham}, {Ramos Almeida}, {Pereira-Santaella}, {Audibert}, {Bellocchi}, {Hicks}, {Labiano}, {Ricci}, \& {Rigopoulou}}]{Arriba_2023}
{Peralta de Arriba}, L., {Alonso-Herrero}, A., {Garc{\'\i}a-Burillo}, S., {et~al.} 2023, \aap, 675, A58, \dodoi{10.1051/0004-6361/202245408}

\bibitem[{{Podgorn{\'y}} {et~al.}(2024{\natexlab{a}}){Podgorn{\'y}}, {Dov{\v{c}}iak}, \& {Marin}}]{Podgorny_2024b}
{Podgorn{\'y}}, J., {Dov{\v{c}}iak}, M., \& {Marin}, F. 2024{\natexlab{a}}, \mnras, 530, 2608, \dodoi{10.1093/mnras/stae1009}

\bibitem[{{Podgorn{\'y}} {et~al.}(2024{\natexlab{b}}){Podgorn{\'y}}, {Marin}, \& {Dov{\v{c}}iak}}]{Podgorny_2024a}
{Podgorn{\'y}}, J., {Marin}, F., \& {Dov{\v{c}}iak}, M. 2024{\natexlab{b}}, \mnras, 527, 1114, \dodoi{10.1093/mnras/stad3266}

\bibitem[{{Podgorn{\'y}} {et~al.}(2024{\natexlab{c}}){Podgorn{\'y}}, {Svoboda}, {Dov{\v{c}}iak}, {Veledina}, {Poutanen}, {Kaaret}, {Bianchi}, {Ingram}, {Capitanio}, {Datta}, {Egron}, {Krawczynski}, {Matt}, {Muleri}, {Petrucci}, {Russell}, {Steiner}, {Bollemeijer}, {Brigitte}, {Castro Segura}, {Emami}, {Garc{\'\i}a}, {Hu}, {Iacolina}, {Kravtsov}, {Marra}, {Mastroserio}, {Mu{\~n}oz-Darias}, {Nathan}, {Negro}, {Ratheesh}, {Rodriguez Cavero}, {Taverna}, {Tombesi}, {Yang}, {Zhang}, \& {Zhang}}]{Podgorny_2024_swj1727d}
{Podgorn{\'y}}, J., {Svoboda}, J., {Dov{\v{c}}iak}, M., {et~al.} 2024{\natexlab{c}}, \aap, 686, L12, \dodoi{10.1051/0004-6361/202450566}

\bibitem[{{Poutanen} {et~al.}(2018){Poutanen}, {Veledina}, \& {Zdziarski}}]{Poutanen_2018}
{Poutanen}, J., {Veledina}, A., \& {Zdziarski}, A.~A. 2018, \aap, 614, A79, \dodoi{10.1051/0004-6361/201732345}

\bibitem[{{Ramsey} {et~al.}(2022){Ramsey}, {Bongiorno}, {Kolodziejczak}, {Kilaru}, {Alexander}, {Baumgartner}, {Breeding}, {Elsner}, {Le Roy}, {McCracken}, {Mitsuishi}, {O'Dell}, {Pavelitz}, {Ranganathan}, {Sanchez}, {Speegle}, {Thomas}, {Weddendorf}, \& {Weisskopf}}]{Ramsey_2022}
{Ramsey}, B.~D., {Bongiorno}, S.~D., {Kolodziejczak}, J.~J., {et~al.} 2022, Journal of Astronomical Telescopes, Instruments, and Systems, 8, 024003, \dodoi{10.1117/1.JATIS.8.2.024003}

\bibitem[{{Remillard} \& {McClintock}(2006)}]{Remillard_2006}
{Remillard}, R.~A., \& {McClintock}, J.~E. 2006, \araa, 44, 49, \dodoi{10.1146/annurev.astro.44.051905.092532}

\bibitem[{Santangelo {et~al.}(2024)Santangelo, Zhang, Feroci, Hernanz, Lu, \& Xu}]{Santangelo2024}
Santangelo, A., Zhang, S.-N., Feroci, M., {et~al.} 2024, The Enhanced X-ray Timing and Polarimetry Mission: eXTP, ed. C.~Bambi \& A.~Santangelo (Singapore: Springer Nature Singapore), 1201--1229, \dodoi{10.1007/978-981-19-6960-7_34}

\bibitem[{{Schnittman} \& {Krolik}(2010)}]{sh2010ApJ...712..908S}
{Schnittman}, J.~D., \& {Krolik}, J.~H. 2010, \apj, 712, 908, \dodoi{10.1088/0004-637X/712/2/908}

\bibitem[{{Shin} {et~al.}(2010){Shin}, {Ostriker}, \& {Ciotti}}]{Shin_2010}
{Shin}, M.-S., {Ostriker}, J.~P., \& {Ciotti}, L. 2010, \apj, 711, 268, \dodoi{10.1088/0004-637X/711/1/268}

\bibitem[{{Soffitta} {et~al.}(2021){Soffitta}, {Baldini}, {Bellazzini}, {Costa}, {Latronico}, {Muleri}, {Del Monte}, {Fabiani}, {Minuti}, {Pinchera}, {Sgro'}, {Spandre}, {Trois}, {Amici}, {Andersson}, {Attina'}, {Bachetti}, {Barbanera}, {Borotto}, {Brez}, {Brienza}, {Caporale}, {Cardelli}, {Carpentiero}, {Castellano}, {Castronuovo}, {Cavalli}, {Cavazzuti}, {Ceccanti}, {Centrone}, {Ciprini}, {Citraro}, {D'Amico}, {D'Alba}, {Di Cosimo}, {Di Lalla}, {Di Marco}, {Di Persio}, {Donnarumma}, {Evangelista}, {Ferrazzoli}, {Hayato}, {Kitaguchi}, {La Monaca}, {Lefevre}, {Loffredo}, {Lorenzi}, {Lucchesi}, {Magazzu}, {Maldera}, {Manfreda}, {Mangraviti}, {Marengo}, {Matt}, {Mereu}, {Morbidini}, {Mosti}, {Nakano}, {Nasimi}, {Negri}, {Nenonen}, {Nuti}, {Orsini}, {Perri}, {Pesce-Rollins}, {Piazzolla}, {Pilia}, {Profeti}, {Puccetti}, {Rankin}, {Ratheesh}, {Rubini}, {Santoli}, {Sarra}, {Scalise}, {Sciortino}, {Tamagawa}, {Tardiola}, {Tobia}, {Vimercati}, \& {Xie}}]{Soffitta_2021}
{Soffitta}, P., {Baldini}, L., {Bellazzini}, R., {et~al.} 2021, \aj, 162, 208, \dodoi{10.3847/1538-3881/ac19b0}

\bibitem[{{Tagliacozzo} {et~al.}(2023){Tagliacozzo}, {Marinucci}, {Ursini}, {Matt}, {Bianchi}, {Baldini}, {Barnouin}, {Cavero Rodriguez}, {De Rosa}, {Di Gesu}, {Dov{\v{c}}iak}, {Harper}, {Ingram}, {Karas}, {Kim}, {Krawczynski}, {Madejski}, {Marin}, {Middei}, {Marshall}, {Muleri}, {Panagiotou}, {Petrucci}, {Podgorny}, {Poutanen}, {Puccetti}, {Soffitta}, {Tombesi}, {Veledina}, {Zhang}, {Agudo}, {Antonelli}, {Bachetti}, {Baumgartner}, {Bellazzini}, {Bongiorno}, {Bonino}, {Brez}, {Bucciantini}, {Capitanio}, {Castellano}, {Cavazzuti}, {Chen}, {Ciprini}, {Costa}, {Del Monte}, {Di Lalla}, {Di Marco}, {Donnarumma}, {Doroshenko}, {Ehlert}, {Enoto}, {Evangelista}, {Fabiani}, {Ferrazzoli}, {Garcia}, {Gunji}, {Heyl}, {Iwakiri}, {Jorstad}, {Kaaret}, {Kislat}, {Kitaguchi}, {Kolodziejczak}, {La Monaca}, {Latronico}, {Liodakis}, {Maldera}, {Manfreda}, {Marscher}, {Massaro}, {Mitsuishi}, {Mizuno}, {Negro}, {Ng}, {O'Dell}, {Omodei}, {Oppedisano}, {Papitto}, {Pavlov}, {Peirson}, {Perri}, {Pesce-Rollins}, {Pilia}, {Possenti},
  {Ramsey}, {Rankin}, {Ratheesh}, {Roberts}, {Romani}, {Sgr{\`o}}, {Slane}, {Spandre}, {Swartz}, {Tamagawa}, {Tavecchio}, {Taverna}, {Tawara}, {Tennant}, {Thomas}, {Trois}, {Tsygankov}, {Turolla}, {Vink}, {Weisskopf}, {Wu}, {Xie}, \& {Zane}}]{Tagliacozzo_2023}
{Tagliacozzo}, D., {Marinucci}, A., {Ursini}, F., {et~al.} 2023, \mnras, 525, 4735, \dodoi{10.1093/mnras/stad2627}

\bibitem[{{Tanimoto} {et~al.}(2023){Tanimoto}, {Wada}, {Kudoh}, {Odaka}, {Uematsu}, \& {Ogawa}}]{Tanimoto_2023}
{Tanimoto}, A., {Wada}, K., {Kudoh}, Y., {et~al.} 2023, \apj, 958, 150, \dodoi{10.3847/1538-4357/ad06ac}

\bibitem[{{Terashima} {et~al.}(2002){Terashima}, {Iyomoto}, {Ho}, \& {Ptak}}]{Terashima_2002}
{Terashima}, Y., {Iyomoto}, N., {Ho}, L.~C., \& {Ptak}, A.~F. 2002, \apjs, 139, 1, \dodoi{10.1086/324373}

\bibitem[{{Ulvestad} \& {Wilson}(1984)}]{Ulvestad1984}
{Ulvestad}, J.~S., \& {Wilson}, A.~S. 1984, \apj, 285, 439, \dodoi{10.1086/162520}

\bibitem[{{Ursini} {et~al.}(2019){Ursini}, {Bassani}, {Malizia}, {Bazzano}, {Bird}, {Stephen}, \& {Ubertini}}]{Ursini_2019}
{Ursini}, F., {Bassani}, L., {Malizia}, A., {et~al.} 2019, \aap, 629, A54, \dodoi{10.1051/0004-6361/201936273}

\bibitem[{{Ursini} {et~al.}(2022){Ursini}, {Matt}, {Bianchi}, {Marinucci}, {Dov{\v{c}}iak}, \& {Zhang}}]{Ursini_2022}
{Ursini}, F., {Matt}, G., {Bianchi}, S., {et~al.} 2022, \mnras, 510, 3674, \dodoi{10.1093/mnras/stab3745}

\bibitem[{{Veledina} {et~al.}(2023){Veledina}, {Muleri}, {Dov{\v{c}}iak}, {Poutanen}, {Ratheesh}, {Capitanio}, {Matt}, {Soffitta}, {Tennant}, {Negro}, {Kaaret}, {Costa}, {Ingram}, {Svoboda}, {Krawczynski}, {Bianchi}, {Steiner}, {Garc{\'\i}a}, {Kravtsov}, {Nitindala}, {Ewing}, {Mastroserio}, {Marinucci}, {Ursini}, {Tombesi}, {Tsygankov}, {Yang}, {Weisskopf}, {Trushkin}, {Egron}, {Iacolina}, {Pilia}, {Marra}, {Miku{\v{s}}incov{\'a}}, {Nathan}, {Parra}, {Petrucci}, {Podgorn{\'y}}, {Tugliani}, {Zane}, {Zhang}, {Agudo}, {Antonelli}, {Bachetti}, {Baldini}, {Baumgartner}, {Bellazzini}, {Bongiorno}, {Bonino}, {Brez}, {Bucciantini}, {Castellano}, {Cavazzuti}, {Chen}, {Ciprini}, {De Rosa}, {Del Monte}, {Di Gesu}, {Di Lalla}, {Di Marco}, {Donnarumma}, {Doroshenko}, {Ehlert}, {Enoto}, {Evangelista}, {Fabiani}, {Ferrazzoli}, {Gunji}, {Hayashida}, {Heyl}, {Iwakiri}, {Jorstad}, {Karas}, {Kislat}, {Kitaguchi}, {Kolodziejczak}, {La Monaca}, {Latronico}, {Liodakis}, {Maldera}, {Manfreda}, {Marin}, {Marscher}, {Marshall},
  {Massaro}, {Mitsuishi}, {Mizuno}, {Ng}, {O'Dell}, {Omodei}, {Oppedisano}, {Papitto}, {Pavlov}, {Peirson}, {Perri}, {Pesce-Rollins}, {Possenti}, {Puccetti}, {Ramsey}, {Rankin}, {Roberts}, {Romani}, {Sgr{\`o}}, {Slane}, {Spandre}, {Swartz}, {Tamagawa}, {Tavecchio}, {Taverna}, {Tawara}, {Thomas}, {Trois}, {Turolla}, {Vink}, {Wu}, \& {Xie}}]{Veledina_2023ApJ_Swj1727a}
{Veledina}, A., {Muleri}, F., {Dov{\v{c}}iak}, M., {et~al.} 2023, \apjl, 958, L16, \dodoi{10.3847/2041-8213/ad0781}

\bibitem[{{Weisskopf} {et~al.}(2022){Weisskopf}, {Soffitta}, {Baldini}, {Ramsey}, {O'Dell}, {Romani}, {Matt}, {Deininger}, {Baumgartner}, {Bellazzini}, {Costa}, {Kolodziejczak}, {Latronico}, {Marshall}, {Muleri}, {Bongiorno}, {Tennant}, {Bucciantini}, {Dovciak}, {Marin}, {Marscher}, {Poutanen}, {Slane}, {Turolla}, {Kalinowski}, {Di Marco}, {Fabiani}, {Minuti}, {La Monaca}, {Pinchera}, {Rankin}, {Sgro'}, {Trois}, {Xie}, {Alexander}, {Allen}, {Amici}, {Andersen}, {Antonelli}, {Antoniak}, {Attin{\`a}}, {Barbanera}, {Bachetti}, {Baggett}, {Bladt}, {Brez}, {Bonino}, {Boree}, {Borotto}, {Breeding}, {Brienza}, {Bygott}, {Caporale}, {Cardelli}, {Carpentiero}, {Castellano}, {Castronuovo}, {Cavalli}, {Cavazzuti}, {Ceccanti}, {Centrone}, {Citraro}, {D'Amico}, {D'Alba}, {Di Gesu}, {Del Monte}, {Dietz}, {Di Lalla}, {Persio}, {Dolan}, {Donnarumma}, {Evangelista}, {Ferrant}, {Ferrazzoli}, {Ferrie}, {Footdale}, {Forsyth}, {Foster}, {Garelick}, {Gunji}, {Gurnee}, {Head}, {Hibbard}, {Johnson}, {Kelly}, {Kilaru}, {Lefevre},
  {Roy}, {Loffredo}, {Lorenzi}, {Lucchesi}, {Maddox}, {Magazzu}, {Maldera}, {Manfreda}, {Mangraviti}, {Marengo}, {Marrocchesi}, {Massaro}, {Mauger}, {McCracken}, {McEachen}, {Mize}, {Mereu}, {Mitchell}, {Mitsuishi}, {Morbidini}, {Mosti}, {Nasimi}, {Negri}, {Negro}, {Nguyen}, {Nitschke}, {Nuti}, {Onizuka}, {Oppedisano}, {Orsini}, {Osborne}, {Pacheco}, {Paggi}, {Painter}, {Pavelitz}, {Pentz}, {Piazzolla}, {Perri}, {Pesce-Rollins}, {Peterson}, {Pilia}, {Profeti}, {Puccetti}, {Ranganathan}, {Ratheesh}, {Reedy}, {Root}, {Rubini}, {Ruswick}, {Sanchez}, {Sarra}, {Santoli}, {Scalise}, {Sciortino}, {Schroeder}, {Seek}, {Sosdian}, {Spandre}, {Speegle}, {Tamagawa}, {Tardiola}, {Tobia}, {Thomas}, {Valerie}, {Vimercati}, {Walden}, {Weddendorf}, {Wedmore}, {Welch}, {Zanetti}, \& {Zanetti}}]{Weisskopf_2022}
{Weisskopf}, M.~C., {Soffitta}, P., {Baldini}, L., {et~al.} 2022, Journal of Astronomical Telescopes, Instruments, and Systems, 8, 026002, \dodoi{10.1117/1.JATIS.8.2.026002}

\bibitem[{{Zhang} {et~al.}(2019){Zhang}, {Dov{\v{c}}iak}, \& {Bursa}}]{Zhang_2019}
{Zhang}, W., {Dov{\v{c}}iak}, M., \& {Bursa}, M. 2019, \apj, 875, 148, \dodoi{10.3847/1538-4357/ab1261}

\end{thebibliography}
\bibliographystyle{aasjournal}



\end{document}